\documentclass[multphys,vecphys]{svmult}

\usepackage{makeidx}     
\usepackage{graphicx}    
\usepackage{multicol}    
\usepackage[T1]{fontenc}
\usepackage[latin1]{inputenc}
\usepackage{graphicx}
\usepackage{amssymb}

\makeindex             

\def\nuc#1#2{\relax\ifmmode{}^{#1}{\protect\text{#2}}\else${}^{#1}$#2\fi}

\begin{document}
\title*{Angular Momentum Projection and Quadrupole Correlations Effects in 
Atomic Nuclei}
\titlerunning{Angular Momentum Projection and Quadrupole Correlations in Nuclei}
\author{
        J. L. Egido
	\and
        L. M. Robledo
}
\institute{Departamento de F\'\i sica Te\'orica C--XI, Universidad
Aut\'onoma de Madrid, E--28049 Madrid, Spain}

\maketitle
%

\section{Introduction}
%

\label{sec:Introduction}The mean field approach when combined with
effective interactions like Skyrme or Gogny is central to the understanding
of nuclear structure as it provides the right magic numbers and a rather
good description of nuclear properties like masses, radii, etc, see \cite{Bender.review}
and references therein. In this approach nucleons are assumed to
move in orbits created by a common potential and therefore the nuclear
wave function for the ground state can be represented as a Slater
determinant built upon the orbitals occupied by the nucleons. The
common potential is determined from the effective interaction by solving
the Hartree- Fock (HF) equation. On the other hand, most of the nuclei
show, at least in their ground state, the phenomenon of nuclear superconductivity
due to a part of the interaction known as pairing interaction. When
this is the case we have to introduce the concept of quasi-particles
(given by the canonical Bogoliubov transformation) and the mean field
wave function now becomes a product wave function of annihilation
quasi-particle operators. The quasi-particle amplitudes are determined
by solving the Hartree- Fock- Bogoliubov (HFB) equation. 

The nuclear mean field has a strong tendency to show the phenomenon
called ``spontaneous symmetry breaking'' \cite{Mang.75,Ring_Schuck.80,Blaizot.85}
that appears when the HF or HFB wave functions do not respect the
underlying symmetries of the Hamiltonian. In fact this is the case, for example,
for the HFB wave functions as they do not have a definite number of
particles, i.e. they spontaneously break the {}``number of particles''
symmetry. As the atomic nucleus is a finite system the spontaneous
symmetry breaking mechanism is a mere artifact of the mean field approximation
to generate correlations
(although it allows an intuitive understanding of some nuclear structure
effects like rotational bands) contrary to what happens in quantum
field theory where it represents a real effect due to the infinite
number of degrees of freedom. Usually in nuclear physics the spontaneous
symmetry breaking has to do with spatial symmetries like the rotational
or parity symmetries. In the former case it leads to the concept of
``deformed mean field'' where the common potential felt by nucleons
is not rotational invariant (i.e. it is deformed in opposition to
a spherical -rotational invariant- potential) and, as a consequence,
the ground state wave function is not an eigenstate of the angular
momentum operators $J^{2}$ and $J_{z}$ (i.e. the Casimir operators
 of the rotation group). However, the real wave function of the nucleus
is an eigenstate of angular momentum and therefore it is necessary
to go beyond the mean field approximation in order to have the right
quantum numbers. The procedure to restore the symmetry 
is known as the Angular Momentum Projection
(AMP) method \cite{Peierls.57} and relies in the fact that when the
{}``deformed mean field'' wave function (in the following the {}``intrinsic''
wave function) is rotated the corresponding intrinsic mean field energy
remains the same. Therefore, a suitable linear combination of such
rotated {}``intrinsic states'' will recover the angular momentum
quantum numbers of the wave function and at the same time will reduce
the energy (what is worthy from a variational point of view). Usually it is
 said in the literature that the {}``deformed
mean field'' wave functions belong to the {}``intrinsic'' frame
of reference whereas the projected wave functions belong to the
{}``laboratory'' frame of reference. In order to go from the intrinsic
frame to the laboratory one the fluctuations in orientation have to
be added to the intrinsic wave function in exactly the way as it is
done in the Angular Momentum Projection framework.

Angular Momentum Projection has been a goal of nuclear physicist 
for many years and only recently with the new computer facilities 
has become a reality for involved forces. Apart from the traditional motivations 
 a considerable effort has been made, in the last few years, in order
to implement Angular Momentum Projection with realistic effective
interactions. The main reason, apart from the genuine desire of always
having the best theoretical description, is the wealth of new experimental
data in exotic regions of the Nuclide Chart far away from the stability
line and coming from the amazing and very sophisticated experimental
setups that have been assembled in the last few years. In many cases,
the experimental results can not be reproduced or even understood
adequately with a mean field description and therefore, in addition to 
considerations concerning the suitability of such interactions away from 
the stability line, effects beyond mean field have to be explored.
As a result of the AMP calculations it turned out that away of the 
stability line the potential energy surfaces are very soft and that 
shape coexistence is a phenomenon rather common in these exotic regions.
In these cases one is forced to perform, besides the AMP, configuration 
mixing calculations. The most effective way to consider shape mixing is
the Generator Coordinate Method (GCM) in which the relevant coordinates 
(in general the multipole moments) are used to generate the corresponding
wave functions. The mixing coefficients are obtained by solving the 
Hill-Wheeler equation \cite{Hill.53}. The combined AMPGCM with effective
forces is a very powerful method which has allowed, as we will see, 
to understand and to predict many new features.

Obviously there are many more methods apart from the one to be discussed
here that are successful in the description the nuclear structure
phenomena at low energies. We have the traditional shell model (see
\cite{Brown.01} for a recent review and \cite{Caurier.98} for the
state of the art implementations of this method) where first a reduced
set of orbitals is chosen as to be the one playing the most important
role in the physics to be described and then the full diagonalization
of the Hamiltonian in the space of multi particle-hole excitations
coupled to the right quantum numbers is performed. Very good results
are obtained with this method when the effective interaction used
is fine tuned to the set of nuclei to be described and also when the
physics to be described lies within the configuration space chosen.
However, there are several drawbacks for this method: a) The dimension
of the Hamiltonian matrix to be diagonalized dramatically increases
with the size of the configuration space restricting its applicability
to mass numbers smaller than 70 and also to states that do not involve
different major shells at the same time. b) Its success depends upon
a careful fitting of the interaction and therefore it is not suited
for exploratory calculations in new regions of the Nuclide Chart.
c) Finally, it is difficult to recast the results in terms of traditional
concepts based on a mean field picture of the nucleus. 

Another interesting approach is variational approach of the T\"{u}bingen
group \cite{Schmid.01,Schmid.87} where the lab frame wave functions
of the nucleus are constructed by projecting, onto all the preserved
quantum numbers, an unrestricted intrinsic mean field wave function.
The method is sometimes complemented by allowing also multi particle-hole
excitations which are also variationally determined and then the set
of wave functions obtained is used to diagonalize the Hamiltonian.
This method fully shares one of the drawbacks of the shell model approach,
namely the one denoted by b) in the previous paragraph and it also
partially suffers from the drawback a) of the shell model, namely
that not so big configuration spaces can be used in the calculations.
Up to now it has been possible to study nuclei in the Kr (Z=36) region
with this method.

An approach that can be extended to heavy nuclei is the Projected
Shell Model (PSM) of Hara and Sun \cite{Hara.95}. In this method
the Hamiltonian is taken as the one of the Pairing+Quadrupole model
with single particle energies fitted to experimental data. The HFB
ground state and many multiquasiparticle excitations are projected
onto good angular momentum and the Hamiltonian is diagonalized in
the resulting basis. One of the advantages of the method over the
two previous ones is that bigger configuration spaces can be used
(up to three harmonic oscillator major shells) but it shares the deficiency
related to the fitting procedure of the interaction in order to get
a reasonable description of experimental data.

Finally, we have the Monte Carlo Shell Model (MCSM) approach of Otsuka
and collaborators \cite{Otsuka.01}. In this method intrinsic wave
functions are generated stochastically and then projected onto the
right quantum numbers. The resulting configuration is kept if its
projected energy is lower than a given threshold generating in this
way a basis of projected wave functions. The Hamiltonian is diagonalized
in this basis at the end. As in the other approaches, rather limited
configuration spaces can be used (although it is possible to carry
out calculations for nuclei as heavy as  Barium ) and the matrix
elements of the interaction have to be carefully fitted to the region
of interest. 

The four approaches just described share the common problem of the
effective charges that have to be introduced in the calculation of
transition probabilities as a consequence of the limited configuration
spaces.

The main advantage of using beyond mean field approaches with effective
forces over the other approaches mentioned before relies in the universal
character of the forces. The phenomenological effective forces are
supposed to be valid all over the Nuclide Chart and are also supposed
to contain all the ingredients needed to describe  well  low
energy nuclear structure phenomena. In addition, the effectives forces
are defined over the whole configuration space (in principle over
the whole Hilbert space, including even continuum states) and therefore
there is no need to specify which orbitals will play a role when a
new phenomenon has to be described as all of them enter the game and
its role will be determined by the optimization of the energy. As
a bonus, no effective charges are needed in the calculation of transition
probabilities. Another important advantage of the method comes from
the fact that the starting point is always the mean field and therefore
the results are much easier to interpret in terms of familiar quantities.
 Up to now the whole AMPGCM has been performed with the quadrupole moment 
and restricted to the axially symmetric case, which hints to the computational
drawbacks  of the method. In principle one would like to take as many
generator coordinates as possible, however for the nowadays computers
a two dimensional AMPGCM sets the limits of  reasonable calculations. The
reason is that the effective forces are defined in the whole configuration 
space. As a consequence, it is capital for the method to perform reasonably
well to have a good guess of the relevant degrees of freedom to describe
a given phenomenon. Global properties as binding energies, quadrupole moments,
etc, as well as  ground and collective excited states are usually very well 
described by the method. However, a very accurate description of any general excited
states  beyond the AMPGCM in its present form.

In the following we will discuss AMPGCM with effective forces of the Skyrme 
and Gogny type. In Sect. \ref{sec:MF}
the mean field approach will be briefly reviewed and the Skyrme and
Gogny forces introduced. In Sect. \ref{sec:AMP} the technicalities
of AMP will be discussed and some examples will be used to illustrate
the procedure. In Sect. \ref{sec:GCM} configuration mixing with AMP
will be presented an its main outcomes will be discussed with an example.
In Sect. \ref{sec:Results} an account of the most remarkable results
obtained so far with both the Skyrme and Gogny interactions will be
presented and briefly discussed. In Sect. \ref{sec:Outlook} we will
end up with an outlook of the results discussed and we will discuss
further developments of the theory.

\section{Symmetry Breaking Mean Field}

\label{sec:MF}The mean field HFB wave functions are determined in terms
of quasi-particle creation $\alpha_{\mu}^{+}$ and annihilation $\alpha_{\mu}=\left(\alpha_{\mu}^{+}\right)^{+}$
operators (Bogoliubov canonical transformation) \[
\alpha_{\mu}^{+}=\sum_{k}U_{k\mu}c_{k}^{+}+V_{k\mu}c_{k}\]
what are given as linear combinations with the amplitudes $U_{k\mu}$
and $V_{k\mu}$ of convenient creation and annihilation single particle
operators $c_{k}^{+}$and $c_{k}$ (usually creating or annihilating
Harmonic Oscillator eigenstates). The ground state wave function
is the product wave function defined by the condition $\alpha_{\mu}|\varphi\rangle=0$
and given by\[
|\varphi\rangle=\prod_{\mu}\alpha_{\mu}|0\rangle\]
where $|0\rangle$ is the true vacuum wave function and the product
runs over all quasi-particle quantum numbers leading to a non zero
wave function $|\varphi\rangle$. The $U_{k\mu}$ and $V_{k\mu}$
amplitudes are determined by requiring the HFB energy to be a minimum
what leads to the well known HFB equation\begin{equation}
\left(\begin{array}{cc}
h & \Delta\\
-\Delta^{*} & -h^{*}\end{array}\right)\left(\begin{array}{cc}
U & V^{*}\\
V & U^{*}\end{array}\right)=\left(\begin{array}{cc}
U & V^{*}\\
V & U^{*}\end{array}\right)\left(\begin{array}{cc}
E & 0\\
0 & -E\end{array}\right)\label{eq:HFB}\end{equation}
This is a non linear equation as the HFB fields $h$ and $\Delta$
depend on the solution through the density matrix and pairing tensor
(see \cite{Mang.75,Ring_Schuck.80,Blaizot.85,Bender.review} for technical
details on how to solve this equation). As was already mentioned in
Sect. \ref{sec:Introduction} the solution of (\ref{eq:HFB}) does not preserve
in many cases the symmetries of the Hamiltonian and is very usual
to find that it breaks rotational invariance leading to a deformed
matter density distribution which is characterized by its multipole
moments like the quadrupole moments $q_{2\mu}$, the octupole
ones $q_{3\mu}$, the hexadecapole ones $q_{4\mu}$, etc.  In order to better 
characterize
the symmetry breaking solution it turns out to be convenient to study
the HFB energy in the neighborhood of the self consistent minimum.
In this way we can study whether the solution corresponds to a well
developed minimum or if there are other local minima around at an
energy relatively close the one found indicating thereby that the
mean field solution obtained has got chances of being unstable when
additional correlations are included. The best way to do such an study
is to carry out constrained calculations where the minimum of the
HFB energy is sought but with the constraint that the mean values
of relevant operators take a give value. In the case at hand, it is
customary to constraint in the mean value of the mass quadrupole operator.
The constrained HFB equation is the same as the unconstrained one
except for the substitution $h\rightarrow h-\lambda o$ where $o$
stands for the matrix of the single particle matrix elements of the
constraining operator and $\lambda$ for the chemical potential that
is determined as a Lagrange multiplier to force the solution to satisfy
the imposed constraint. 

For the interaction to be used in the HFB calculation there are several
possible choices but nowadays the most popular options seem to be
the effective density dependent interactions of Skyrme or Gogny type.
Both are very similar in structure the only difference being that
the former is a zero range force contrary to the latter and therefore
it cannot be used for the calculation of the matrix elements entering
into the definition of the pairing field. 

Both the Skyrme \cite{Skyrme.56,Vautherin.72,Engel.75} and Gogny
\cite{Dech_Gogny.80} interactions are two body, density dependent
phenomenological interaction given by the sum of four terms \[
v(1,2)=v_{C}+v_{LS}+v_{DD}+v_{\mathrm{Coul}}.\]
 The Coulomb field $v_{\mathrm{Coul}}=e^{2}/|\vec{r}_{1}-\vec{r}_{2}|$,
the density dependent interaction $v_{DD}=t_{3}(1+P_{\sigma}x_{0})\delta(\vec{r}_{1}-\vec{r}_{2})\rho^{\alpha}((\vec{r}_{1}+\vec{r}_{2})/2)$,
and the two body, non-relativistic spin orbit potential $v_{LS}=iW_{LS}(\vec{\nabla}_{12}\delta(\vec{r}_{1}-\vec{r}_{2})\wedge\vec{\nabla}_{12})(\vec{\sigma}_{1}+\vec{\sigma}_{2})$
are the same or very similar in both interactions (they usually differ
in the power $\alpha$ entering the density dependent term). However,
the central potential $v_{C}$ is different in both interactions being
of zero range in the case of the Skyrme interaction \begin{eqnarray*}
v_{C}(\mathrm{Skyrme}) & = & t_{0}(1+x_{0}P_{\sigma})\delta(\vec{r}_{1}-\vec{r}_{2})\\
 & + & \frac{1}{2}t_{1}(1+x_{1}P_{\sigma})[\vec{k'}_{12}^{2}\delta(\vec{r}_{1}-\vec{r}_{2})+\delta(\vec{r}_{1}-\vec{r}_{2})\vec{k}_{12}^{2}]\\
 & + & t_{2}(1+x_{2}P_{\sigma})\vec{k'}_{12}^{2}\delta(\vec{r}_{1}-\vec{r}_{2})\vec{k}_{12}^{2}\end{eqnarray*}
whereas it is of \textbf{finite range} in the case of the Gogny interaction
\[
v_{C}(\mathrm{Gogny})=\sum_{i=1,2}e^{-(\vec{r}_{1}-\vec{r}_{2})^{2}/\mu_{i}^{2}}\left(W_{i}+B_{i}P_{\sigma}-H_{i}P_{\tau}-M_{i}P_{\sigma}P_{\tau}\right).\]
 Both contain the usual combinations of spin and isospin projector
operators $P_{\sigma}$ and $P_{\tau}$. In the case of Gogny the finite
range is modeled by the sum of two Gaussian of different ranges. 
The finite range was introduced to prevent the ultraviolet catastrophe
that shows up in the evaluation of the pairing potential. As a consequence,
it can be used for both the particle-hole and particle-particle channel
of the HFB equations. This is not the case for the Skyrme interaction
and extra pairing interactions with their corresponding parameters
are usually introduced to deal with the particle-particle channel.
Concerning the parametrizations of the force, in the case of Skyrme
there are over a hundred parameterizations (like SIII\cite{SkyrIII},
SkM\cite{SkM}, SkM{*} \cite{SkM*} or the Lyon forces \cite{SLy4,SLy6})
that more or less yield the same bulk nuclear properties at the saturation
density but differ in other aspects like their behavior in infinite
nuclear and/or neutron matter or their performance to describe the
collective response of the nucleus. For the Gogny interaction there
are very few parametrizations and in most calculations only the
so called D1S is used. It was adjusted
back in 1984 \cite{Berger.84} to reproduce nuclear matter properties
as well as bulk properties of some selected finite nuclei. As this
parameterization has performed rather well in all the cases where
it has been applied it has been kept over the years. However, taking
into account the poor performance of Gogny for neutron matter a new
parameterization was proposed \cite{GognyD1P} but it has not thoroughly
been tested up to now. A recent review on the mean field properties
of effective density dependent interactions like the ones discussed
here can be found in \cite{Bender.review}.

Since we are focusing in this lecture on the rotational symmetry breaking
mechanism, which is mainly characterized by the mass quadrupole moment
of the density distribution, the starting point of the calculations 
(both with the Skyrme or Gogny interaction) is  a
constrained HFB calculation with the mass quadrupole components as
constrained quantities. As the resulting wave functions will be used
in the context of Angular Momentum Projection, which represents a
very tough computational problem, the calculations will be restricted
to axially symmetric, parity conserving configurations denoted by
$|\varphi(q_{20})\rangle$ (i.e. by construction only the $\mu=0$
component of the quadrupole tensor can be different from zero). These
HFB wave functions are obtained as a solution of the corresponding
HFB equation with the constraint in the mean value of the $\mu=0$
mass quadrupole operator $\left\langle \varphi(q_{20})\right|z^{2}-1/2(x^{2}+y^{2})\left|\varphi(q_{20})\right\rangle =q_{20}$.
Also in both kind of calculations it is customary to neglect the contributions
of the Coulomb field to the exchange and pairing potentials due to
the high computational cost associated with the calculation of those
fields (see \cite{Anguiano.01} for a thorough discussion of these
contributions). In the case of the Gogny force calculations we have
added the exchange Coulomb energy evaluated in the Slater approximation
at the end of the calculation in a perturbative fashion.

\begin{figure}
\begin{center}\includegraphics[%
  clip,
  width=0.95\textwidth,
  keepaspectratio]{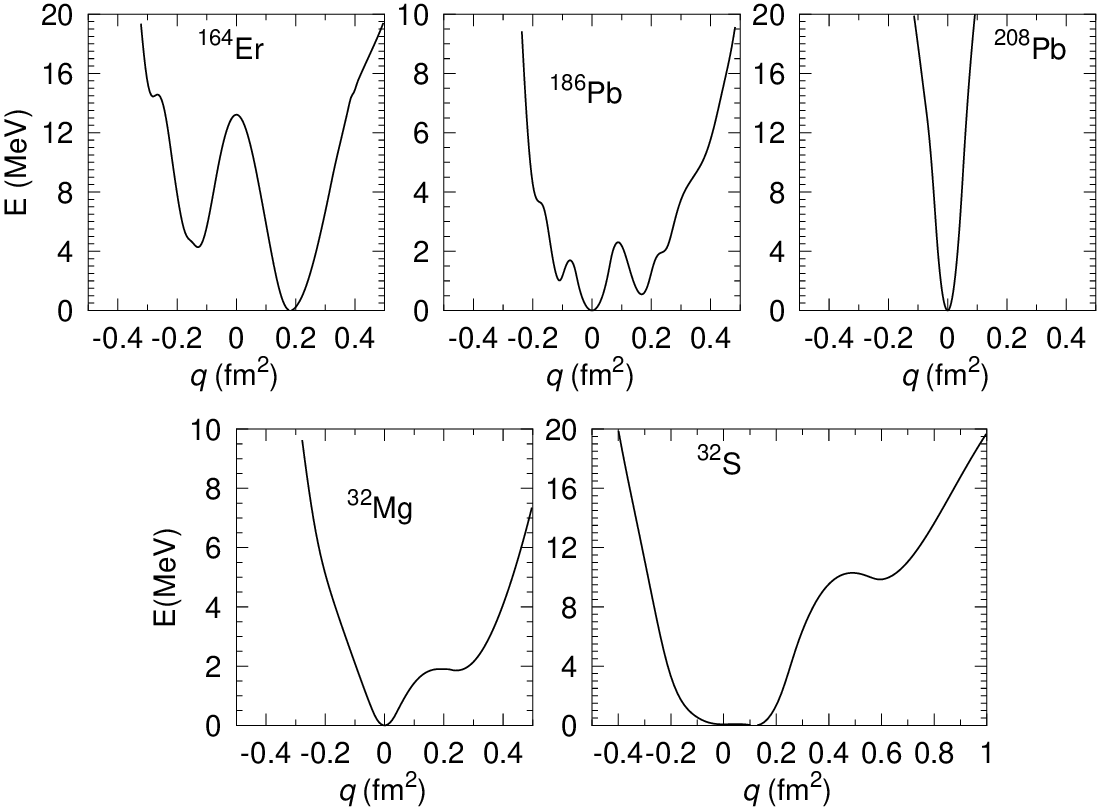}\end{center}

\caption{\label{fig:HFBE}HFB energies as a function of the quadrupole deformation
parameter $q=q_{20}/A^{5/3}$ for five relevant nuclei. Energies are
relative to the respective absolute minimum}
\end{figure}

Usually the HFB equation is solved for the Skyrme interaction in coordinate
space representation by introducing a Cartesian mesh whereas in the
case of the Gogny force a Harmonic Oscillator (HO) basis is used including
up to $N_{0}$ complete shells. The HO length parameters are chosen
to be equal in order to preserve the rotational invariance of the
basis which is a very important requirement for the subsequent Angular
Momentum Projection calculations \cite{Robledo.94}.

As an example of the kind of results it is possible to obtain we have
plotted in Fig. \ref{fig:HFBE} the HFB energies computed with the
Gogny force as a function of $q=q_{20}/A^{5/3}$ (the variable $q$
is defined to be independent of mass number $A$, like the deformation
parameter $\beta$) for five representative nuclei, namely $^{32}$Mg,
$^{32}$S, $^{164}$Er, $^{186}$Pb and $^{208}$Pb. These nuclei
have been chosen as they represent typical cases: In the doubly magic
$^{208}$Pb there is no spontaneous symmetry breaking and the energy
has a minimum at $q=0$ with a very deep and stiff well. In the nucleus
$^{164}$Er we have two minima one prolate ($q>0$) which is the absolute
minimum and the other oblate ($q<0$) separated by a rather high barrier.
In the nucleus $^{32}$S we have an spherical minimum but at a given
deformation an excited local minimum appears corresponding to a super-deformed
configuration (obviously there are better examples of super-deformed
states both in the rare earth region as well as in the Hg region;
we have chosen this particular example as is the only one where subsequent
AMP calculations have been carried out). In the case of the neutron
deficient magic $^{186}$Pb we have three minima one spherical (the
absolute minimum) and two excited minima, one prolate and the other
oblate, which have energies very close to the one of the ground state
(shape coexistence). Finally, in the neutron rich $N=20$ magic nucleus
$^{32}$Mg we obtain a spherical minimum ($q=0$) and a shoulder at
around 1 MeV excitation energy. Similar results are obtained for the
Skyrme interaction as we will see in Sect. \ref{sec:Results}.

\section{Angular Momentum Projection}

\label{sec:AMP}In the following subsections we will introduce all
the technology to carry out angular momentum projection without entering
into much details as most of them can be found in the literature.
However, we will devote some more attention to the issue of how to
deal with density dependent interactions.

\subsection{The Projector Operator}

\label{sub:Projector}Wave functions $\mid\Psi_{IM}\rangle$ eigenstates
of $J^{2}$ and $J_{z}$ with eigenvalues $\hbar^{2}I(I+1)$ and $\hbar M$
respectively can be built out of a given deformed mean field state
$\mid\varphi\rangle$ by applying the angular momentum projector

\begin{eqnarray}
\mid\Psi_{IM}\rangle & = & \sum_{K}g_{K}^{I}\hat{P}_{MK}^{I}\mid\varphi\rangle.\label{PROJWF}\end{eqnarray}
 The angular momentum projection operator $\hat{P}_{MK}^{I}$ is given
by \cite{Peierls.57}

\begin{eqnarray}
\hat{P}_{MK}^{I} & = & \frac{2I+1}{8\pi^{2}}\int d\Omega\mathcal{D}_{MK}^{I*}(\Omega)\hat{R}(\Omega)\label{AMProj}\end{eqnarray}
where $\Omega$ represents the set of the three Euler angles $\left(\alpha,\beta,\gamma\right)$,
$\mathcal{D}_{MK}^{I}(\Omega)$ are the well known Wigner functions
\cite{Varsha.88} and $\hat{R}(\Omega)=e^{-i\alpha\hat{J}_{z}}e^{-i\beta\hat{J}_{y}}e^{-i\gamma\hat{J}_{z}}$
is the rotation operator. That $\mid\Psi_{IM}\rangle$ is an angular
momentum eigenstate is very easy to check. Applying the rotation operator
$\hat{R}(\Omega)$ to it we obtain\[
\hat{R}(\Omega)\mid\Psi_{IM}\rangle=\frac{2I+1}{8\pi^{2}}\sum_{K}g_{K}^{I}\int d\Omega'\mathcal{D}_{MK}^{I*}(\Omega')\hat{R}(\Omega+\Omega')\mid\varphi\rangle\,.\]
Now the integral on the three Euler angles is rewritten as\[
\int d\Omega'\mathcal{D}_{MK}^{I*}(\Omega'-\Omega)\hat{R}(\Omega')\mid\varphi\rangle=\sum_{L}\mathcal{D}_{ML}^{I}(\Omega)\int d\Omega'\mathcal{D}_{LK}^{I*}(\Omega')\hat{R}(\Omega')\mid\varphi\rangle\]
which shows that \[
\hat{R}(\Omega)\mid\Psi_{IM}\rangle=\sum_{L}\mathcal{D}_{ML}^{I}(\Omega)\mid\Psi_{IL}\rangle\]
The quantities $g_{K}^{I}$ are arbitrary at this point and are usually
determined in a variational sense to yield a minimum of the projected
energy. The resulting equation will be discussed in the next subsection.
Other relevant properties of the projection operator can be easily
deduced from the following representation of the projector \cite{Corbett.71}
\begin{equation}
P_{MK}^{I}=\sum_{\alpha}|\alpha IM\rangle\langle\alpha IK|\,.\label{eq:PIMKCR}\end{equation}
Using this representation is very easy to derive the property \begin{equation}
\left(P_{MK}^{I}\right)^{+}P_{M'K'}^{I'}=\delta_{II'}\delta_{MM'}P_{KK'}^{I}\label{eq:PPP}\end{equation}
as well as\begin{equation}
\left(P_{MK}^{I}\right)^{+}=P_{KM}^{I}\,.\label{eq:PPLUS}\end{equation}

\begin{figure}
\begin{center}\includegraphics[%
  clip,
  width=1.0\textwidth]{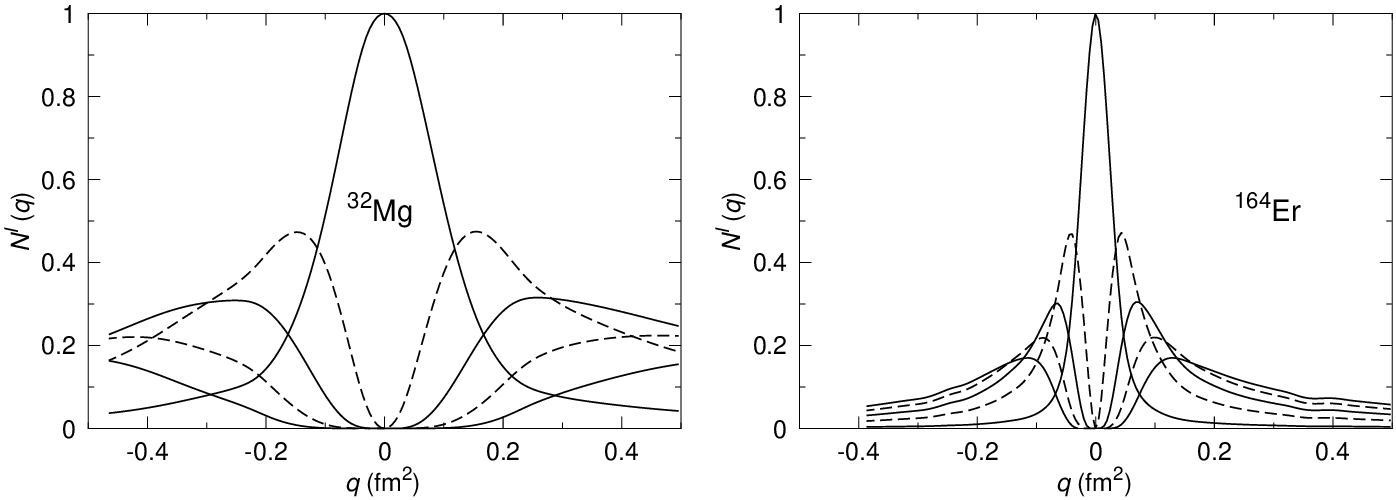}\end{center}

\caption{\label{fig:Norm}Projected norms $\mathcal{N}^{I}(q)$ for $I=0$,
2, 4, 6 and 8 (dashed lines correspond to $I=2$ and 6 ) plotted as
a function of the quadrupole deformation parameter $q=q_{20}/A^{5/3}$
for $^{32}$Mg and $^{164}$Er}
\end{figure}

An interesting quantity is the overlap between the intrinsic wave
function $|\varphi\rangle$ and the projected wave functions $\mid\Psi_{IM}\rangle$
that gives the probability amplitude of finding the angular momentum
$I$ and third component $M$ in the intrinsic wave function. In Fig.
\ref{fig:Norm} we have plotted such quantities as a function of the
quadrupole deformation for the axially symmetric HFB solutions obtained
for the nuclei $^{32}$Mg and $^{164}$Er that were discussed in Sect.
\ref{sec:MF}. As the intrinsic wave functions are axially symmetric
the only meaningful quantity in this case is $\mathcal{N}^{I}(q_{20})=\langle\varphi(q_{20})|P_{00}^{I}|\varphi(q_{20})\rangle$
that is the quantity plotted. We observe that the spherical solution
($q_{20}=0$) only has $I=0$ component or in other words the HFB
solution corresponding to $q_{20}=0$ is already an eigenstate of
angular momentum with $I=0$. For increasing deformation the $I=0$
probability decreases whereas the other components increase up to
a given point and from there on they decrease. Finally, for large
deformations all the components are of the same order of magnitude
implying a strong spreading of angular momentum for the intrinsic
state.

Finally, it has to be mentioned that there are other representations
of the projector $P_{MK}^{I}$ as the one of \cite{Kerman.77} that
use the rotation operator in the form $\hat{R}(\kappa)=\exp(-i\kappa\vec{n}\vec{J})$.

\subsection{The Projected Energy}

\label{sub:PROJE}The projected energy $E^{I}$ is simply given by

\begin{equation}
E^{I}=\frac{\langle\Psi_{IM}\mid\hat{H}\mid\Psi_{IM}\rangle}{\langle\Psi_{IM}\mid\Psi_{IM}\rangle}
\label{PROJENER}\end{equation}
and the fact that this quantity is independent of $M$ is a direct
consequence of the invariance of the Hamiltonian under rotations. 

Up to now nothing has been said about the $g_{K}^{I}$ coefficients
as the geometrical properties of $\mid\Psi_{IM}\rangle$ do not depend
on them. However, we can exploit this degree of freedom in a variational
sense to minimize the projected energy with respect to those parameters.
The resulting equation reads \[
\sum_{K'}g_{K'}^{I}\left(\langle\varphi|\hat{H}P_{KK'}^{I}|\varphi\rangle-E^{I}\langle\varphi|P_{KK'}^{I}|\varphi\rangle\right)=0\,.\]

To compute $\langle\varphi|\hat{H}P_{KK'}^{I}|\varphi\rangle$ we
have to perform a three dimensional integral over the Euler angles
of an integrand that involves the well known Wigner functions and
the Hamiltonian overlap $\langle\varphi|\hat{H}\hat{R}(\Omega)|\varphi\rangle$.
The latter quantity can be straightforwardly evaluated by means of
the extended Wick theorem \cite{Balian.69} and the sign problem of
the norm dealt with as suggested in \cite{Neergard.83}, but it turns
out that its evaluation for the case at hand is a very intensive computational
task that has to be repeated for all the Euler angle mesh points used
to evaluate the three dimensional integral numerically. For present
single CPU computers this is a task that demands of the order of tens
of hours to complete. However, for the Skyrme interaction there has
been an attempt \cite{Bonche.91} to carry out the full calculation
for low spins by restricting the number of mesh points in the $\alpha$
and $\gamma$ integrations and this is a path worth to be explored
in the future. For the other approaches mentioned in the introduction
the computational cost is much smaller due to the restricted configuration
spaces and therefore the full projection is routinely carried out.
Fortunately, in many cases the restriction to axially symmetric intrinsic
wave functions seems to be a sound approximation \cite{Hagino.03} that reduces the
computational burden by almost two orders of magnitude. 
For even-even nuclei when $|\varphi\rangle$
is axially symmetric it satisfies $J_{z}|\varphi\rangle=0$ and therefore
$\langle\varphi|\hat{H}\hat{R}(\Omega)|\varphi\rangle$ reduces to
$\langle\varphi|\hat{H}e^{-i\beta J_{y}}|\varphi\rangle$ which is
independent of $\alpha$ and $\gamma$. As a consequence the integrals
in $\alpha$ and $\gamma$ are trivial and yields $(2\pi)^{2}\delta_{K,0}\delta_{K',0}$.
With this restriction we get for the projected energy\begin{equation}
E^{I}=\frac{\int_{0}^{\pi}d\beta\sin(\beta)d_{00}^{I}(\beta)\langle\varphi|
\hat{H}e^{-i\beta J_{y}}|\varphi\rangle}{\int_{0}^{\pi}d\beta\sin(\beta)
d_{00}^{I}(\beta)\langle\varphi|e^{-i\beta J_{y}}|\varphi\rangle}=
\frac{\int_{0}^{\pi}d\beta\sin(\beta)d_{00}^{I}(\beta)h(\beta)n(\beta)}
{\int_{0}^{\pi}d\beta\sin(\beta)d_{00}^{I}(\beta)n(\beta)}\label{eq:EIAX}
\end{equation}
with $n(\beta)=\langle\varphi|e^{-i\beta J_{y}}|\varphi\rangle$ and
$h(\beta)=\langle\varphi|He^{-i\beta J_{y}}|\varphi\rangle$.
Using as self-consistent symmetry for the intrinsic wave function
the signature (essentially the inversion of the $x$ axis, and given
by the operator $\Pi e^{-i\pi J_{x}}$) it is possible to reduce the
integration interval to $[0,\pi/2]$ and to show that for even-even
nuclei the projected energy is only defined for even values of the
angular momentum $I$ when the intrinsic state is reflection symmetric.
When the intrinsic state is not reflection symmetric the reduction
of the interval of integration implies a projection onto good parity
$\pi$ in addition to the angular momentum projection and the rule
$(-1)^{I}=\pi$ is obtained.

\subsection{How to Deal with the Density Dependent Term of the Interactions}

\label{sub:DD}When dealing with density dependent (DD) forces we
face the problem that the DD term is only defined at the mean field
level: In the evaluation of the expectation value of the Hamiltonian
$\langle\varphi|\hat{H}|\varphi\rangle$ the DD interaction is given
in terms of $\rho(\vec{R)}=\langle\varphi|\hat{\rho}(\vec{R)}|\varphi\rangle$
which is evaluated as an expectation value with the \textbf{same}
wave function used in the evaluation of the energy. However, in the
calculation of the projected energies we need the Hamiltonian overlap
$\langle\varphi|\hat{H}\hat{R}(\Omega)|\varphi\rangle$ and, as the
DD term has a phenomenological origin, it is not clear which density
should be used in the DD term of the interaction for the evaluation
of the above mentioned overlap. The same problem arises with the overlap
needed in the Generator Coordinate Method (GCM), namely $\langle\varphi_{0}|\hat{H}|\varphi_{1}\rangle$.
Several \textbf{prescriptions} have been proposed to cope with this
problem: in one \cite{Duguet.03B}, inspired by a generalized Brueckner
expansion, the density is replaced by the linear
combination $\frac{1}{2}(\langle\varphi|\hat{\rho}(\vec{R)}|\varphi\rangle+\langle\varphi|\hat{\rho}(\vec{R})\hat{R}(\Omega)|\varphi\rangle)$;
in other called the {}``mixed density'' prescription the quantity
$\langle\varphi|\hat{\rho}(\vec{R})\hat{R}(\Omega)|\varphi\rangle/\langle\varphi|\hat{R}(\Omega)|\varphi\rangle$
is proposed \cite{Bonche.90} (see also \cite{Valor.00} for a thorough
discussion in the context of Particle Number Projection) and it is
inspired by the way the Hamiltonian overlap is computed with the aid
of the extended Wick theorem. There are some more prescriptions, like
the one proposing a rotational invariant density dependent term, but
we will not comment them here. In any case, as in most of the prescriptions
the DD term breaks rotational invariance and even hermiticity, 
it is very important to show that the prescriptions are consistent
with general properties of the projected energy like being a real
quantity and that the projected energy is independent of $M$, i.e., 
Eq.~(\ref{PROJENER}) is satisfied. In \cite{Rayner.2002b} has been shown that 
the {}``mixed density'' prescription 
$\langle\varphi|\hat{\rho}(\vec{R})\hat{R}(\Omega)|
\varphi\rangle/\langle\varphi|\hat{R}(\Omega)|\varphi\rangle$
satisfies the two previous requirements as it also does the other
prescription. At this point it might seem paradoxical to have a Hamiltonian
which is not rotational invariant but this apparent paradox can be
solved if density dependent Hamiltonian are thought not as genuine
Hamiltonian but rather as devices (in the spirit of the Density Functional
Theory) to get an elaborated energy functional of the density.

\subsection{Transition Probabilities and Spectroscopic Factors}

\label{sub:TransProb}Transition probabilities are the physical quantities
that have most sensitivity to the approximations made to the wave
functions. For instance, transition probabilities have selection rules
that can not be reproduced unless the wave functions used are eigenstates
of the angular momentum Casimir operators $J^{2}$ and $J_{z}$. In
this section we will present the way they are computed when the wave
functions used come from an angular momentum projected intrinsic state
and specialized formulas for the case of an axially symmetric intrinsic
state will be presented. 

The only property we will need is the transformation law of the multipole
operators $\hat{Q}_{\lambda\mu}$ under rotations

\begin{equation}
\hat{R}(\Omega)\hat{Q}_{\lambda\mu}\hat{R}^{\dagger}(\Omega)=\sum_{\mu^{'}}\mathcal{D}_{\mu^{'}\mu}^{\lambda}(\Omega)\hat{Q}_{\lambda\mu^{'}}\end{equation}
Using the well known result for the product of two Wigner functions
\cite{Varsha.88} as well as the definition of (\ref{AMProj}) for
the angular momentum projection operator and the property

\begin{equation}
\hat{P}_{MK}^{I}\hat{P}_{K'M'}^{I'}=\delta_{I,I'}\delta_{K,K'}\hat{P}_{MM'}^{I}\end{equation}
 we obtain after some algebra the result

\begin{eqnarray}
\hat{P}_{K_{f}M_{f}}^{I_{f}}\hat{Q}_{\lambda\mu}\hat{P}_{M_{i}K_{i}}^{I_{i}} & = & \langle I_{i}M_{i}\lambda\mu\mid I_{f}M_{f}\rangle\nonumber \\
 & \times & \sum_{\nu\mu^{'}}(-)^{\mu^{'}-\mu}\langle I_{i}\nu\lambda\mu^{'}\mid I_{f}K_{f}\rangle\hat{Q}_{\lambda\mu^{'}}\hat{P}_{\nu K_{i}}^{I_{i}}\,.\end{eqnarray}
 With the definition of the projected wave functions (\ref{PROJWF})
and the previous result we obtain \begin{equation}
\langle\Psi_{I_{f}M_{f}}\mid\hat{Q}_{\lambda\mu}\mid\Psi_{I_{i}M_{i}}\rangle=\frac{\langle I_{i}M_{i}\lambda\mu\mid I_{f}M_{f}\rangle}{\sqrt{2I_{f}+1}}\langle I_{f}\mid\mid\hat{Q}_{\lambda}\mid\mid I_{i}\rangle\end{equation}
 with

\begin{eqnarray}
\langle I_{f}\mid\mid\hat{Q}_{\lambda}\mid\mid I_{i}\rangle & = & \frac{(2I_{i}+1)(2I_{f}+1)}{8\pi^{2}}(-)^{I_{i}-\lambda}\sum_{K_{i}K_{f}\nu\mu^{'}}(-)^{K_{f}}g_{K_{f}}^{I_{f}*}g_{K_{i}}^{I_{i}}\nonumber \\
 & \times & \left(\begin{array}{ccc}
I_{i} & \lambda & I_{f}\\
\nu & \mu^{'} & -K_{f}\end{array}\right)\int d\Omega\mathcal{D}_{\nu K_{i}}^{I_{i}*}(\Omega)\langle\varphi_{f}\mid\hat{Q}_{\lambda\mu^{'}}\hat{R}(\Omega)\mid\varphi_{i}\rangle\,.\label{eq:QRED}\end{eqnarray}
The previous derivation only uses the tensor properties of the electric
multipole operator and therefore also applies to the case of the magnetic
multipole operators. 

Taking advantage of the axial symmetry of the intrinsic wave function
as well as the self-consistent symmetry $\Pi e^{-i\pi\hat{J}_{x}}$
we can simplify the above expressions as follows. First we have \begin{equation}
\langle\varphi_{f}\mid\hat{Q}_{\lambda{\mu}^{'}}\hat{R}(\Omega)\mid\varphi_{i}\rangle=e^{i\alpha\mu^{'}}\langle\varphi_{f}\mid\hat{Q}_{\lambda{\mu}^{'}}e^{-i\beta\hat{J}_{y}}\mid\varphi_{i}\rangle\end{equation}
 that leads to \begin{eqnarray}
 & \int d\Omega\mathcal{D}_{QK_{i}}^{I_{i}*}(\Omega)\langle\varphi_{f}\mid\hat{Q}_{\lambda{\mu}^{'}}\hat{R}(\Omega)\mid\varphi_{i}\rangle=4\pi^{2}\delta_{Q-{\mu}^{'}}\delta_{K_{i}0}\nonumber \\
 & \times\int_{0}^{\pi}d\beta\sin(\beta)d_{-{\mu}^{'}0}^{I_{i}*}(\beta)\langle\varphi_{f}\mid\hat{Q}_{\lambda{\mu}^{'}}e^{-i\beta\hat{J}_{y}}\mid\varphi_{f}\rangle\,.\end{eqnarray}
 Applying this result to the expression (\ref{eq:QRED}) we obtain
\begin{eqnarray}
\langle I_{f}\mid\mid\hat{Q}_{\lambda}\mid\mid I_{i}\rangle & = & \frac{(2I_{i}+1)(2I_{f}+1)}{2}(-)^{I_{i}-\lambda}\sum_{\mu^{'}}\left(\begin{array}{ccc}
I_{i} & \lambda & I_{f}\\
-\mu^{'} & \mu^{'} & 0\end{array}\right)\nonumber \\
 & \times & \int_{0}^{\pi}d\beta\sin(\beta)d_{-{\mu}^{'}0}^{I_{i}*}(\beta)\langle\varphi_{f}\mid\hat{Q}_{\lambda{\mu}^{'}}e^{-i\beta\hat{J}_{y}}\mid\varphi_{i}\rangle\nonumber \\
 & = & (2I_{i}+1)(2I_{f}+1)(-)^{I_{i}-\lambda}\frac{1+(-)^{I_{i}}}{2}\sum_{{\mu}^{'}}\left(\begin{array}{ccc}
I_{i} & \lambda & I_{f}\\
-\mu^{'} & \mu^{'} & 0\end{array}\right)\nonumber \\
 & \times & \int_{0}^{\frac{\pi}{2}}d\beta\sin(\beta)d_{-{\mu}^{'}0}^{I_{i}*}(\beta)\langle\varphi_{f}\mid\hat{Q}_{\lambda{\mu}^{'}}e^{-i\beta\hat{J}_{y}}\mid\varphi_{i}\rangle\end{eqnarray}
 where we have reduced, in the last line, the integration interval
to half the original one making use of the self-consistent signature
symmetry of the intrinsic wave function.

\subsection{Variation Before and After Projection}

\label{sub:PBV}Now that we know how to compute the projected quantities
we can take for a given nucleus its mean field ground state and project
it out to obtain the projected energies for different angular momenta.
This procedure is called the Projection After Variation (PAV) method
as the intrinsic state is determined at the mean field level and projected
afterwards. For the nuclei used as examples in Sect. \ref{eq:HFB}
we will obtain for the $I=0$ ground state the same projected energy
as the intrinsic one for all the examples discussed except for $^{164}$Er
where a lowering of 3.07 MeV is obtained. For $I=2$ the projected
energy for those nuclei with an spherical intrinsic state is an indeterminacy
of the type zero divided by zero. For the deformed $^{164}$Er we
will obtain for $I=2$ an excited state which is 139.4 keV higher
than the $I=0$ projected ground state. 

\begin{figure}
\begin{center}\includegraphics[%
  width=0.95\textwidth]{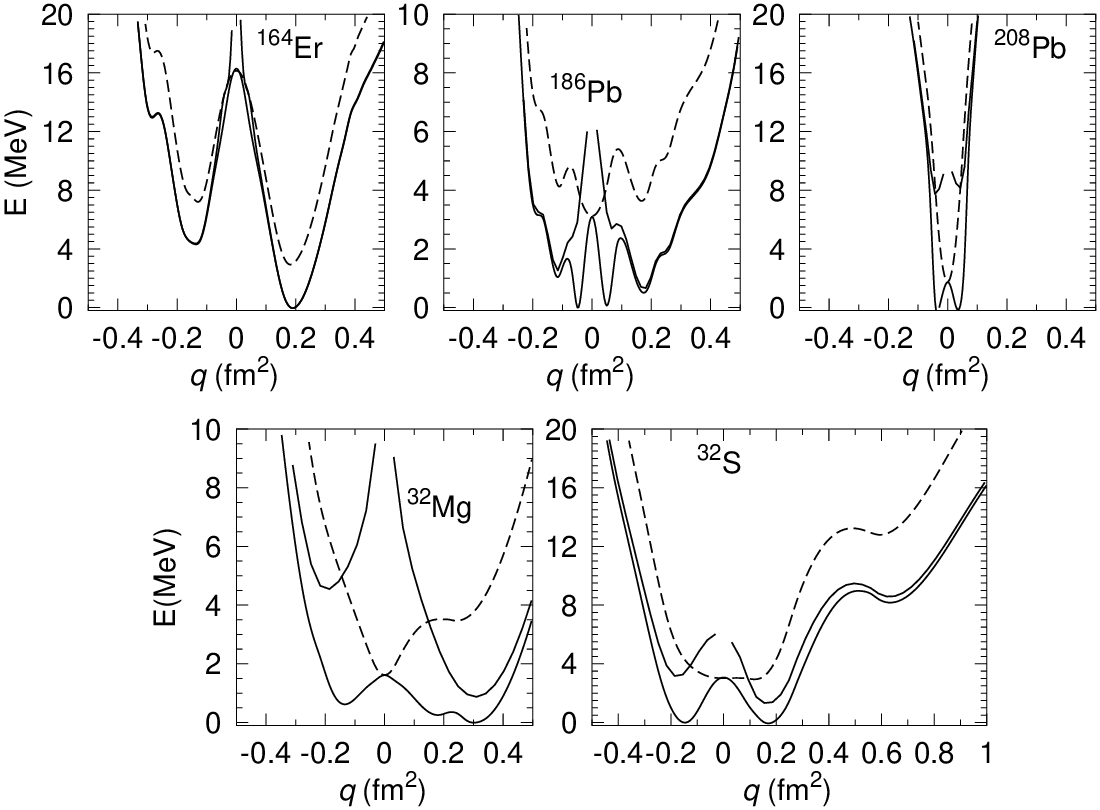}\end{center}

\caption{\label{fig:AMPE}Angular momentum projected energies for $I=0$ and
$I=2$ along with the HFB energy (dashed curves) plotted as a function
of the quadrupole deformation parameter $q=q_{20}/A^{5/3}$ for five
relevant nuclei.}
\end{figure}

As the energy gain due to projection for the ground state of $^{164}$Er
was of the order of a few MeV one may wonder whether this energy gain
can be strong enough for some deformed configurations as to overcome
the mean field energy differences and produce projected energies which
could be lower than the one given by the PAV method. To elucidate
this possibility we have plotted in Fig. \ref{fig:AMPE} the $I=0$
and $I=2$ projected energies along with the HFB results (dashed line)
for all the nuclei considered in Sect. \ref{sec:MF}.

For the well deformed $^{164}$Er nucleus the effect of projection
is just an overall shift of the energy with respect to the HFB curve
except around $q=0$ where for $I=0$ the projected energy has to
coincide with the intrinsic one. In this nucleus the $I=2$ curve
is so close to the $I=0$ one that is almost indistinguishable from
the latter with the scale used for the plot, except around $q=0$
where the intrinsic wave function has an overlap zero with the $I=2$
projected state (see Fig. \ref{fig:Norm} ). 

Not surprisingly the projected energies show minima in all nuclei
considered (except $^{164}$Er) which correspond to deformed intrinsic
configurations~! In addition, the intrinsic configurations corresponding
to the $I=0$ and $I=2$ minima are in the examples presented very
close but they do not share exactly the same intrinsic state. If we
now take as the wave function of the system for each angular momentum
$I$ the one that gives the lower projected energy as a function of
$q$ we will be doing things properly (the variational principle again)
and we will be using a restricted version of the Projection Before
Variation method (PBV). In the PBV method the intrinsic wave function
for each angular momentum is determined by minimizing not the intrinsic
energy but the projected one. The example mentioned above is a restricted
version of the PBV method as we are not exploring, by considering
only wave functions constrained to a given quadrupole deformation,
the whole Hilbert space. However, we can argue that as the quadrupole
moment is the main quantity characterizing the amount of rotational
symmetry breaking, it is the most relevant degree of freedom in terms
of the PBV method and therefore just considering it gives results
very close to the ones of the exact PBV procedure.

At this point it has to be said that except the shell model, all the
techniques mentioned in the introduction aiming to a better description
of the nuclear many body problem are different versions (or views)
of the PBV method. What is different among them is the way the Hilbert
space is searched (fully in the Tubingen approach, by means of multiquasiparticle
excited states in the PSM, stochastically in the MCSM, etc).

Coming back to the discussion of Fig. \ref{fig:AMPE} it is amazing
the large effect caused by projection for the $I=0$ ground state
even in spherical nuclei~! The two light nuclei $^{32}$Mg and $^{32}$S
become well deformed when projection is considered (in good agreement
with the experimental results for $^{32}$Mg). Even the double magic
nucleus $^{208}$Pb becomes slightly deformed\footnote{In fact there are
two deformed minima, one prolate and the other oblate with approximately
the same energy and the same absolute value of the quadrupole moment. In these
cases  as we will see in Section \ref{sec:GCM} one should perform configuration
mixing calculations. The resulting wave function provides an spherical
nucleus and an additional energy lowering of about 1 MeV} and in this way it 
gains 1.8 MeV in correlation energy. This is an important fact (also observed
in other doubly magic nuclei like $^{16}$O, $^{40}$Ca and $^{48}$Ca)
that should be taken into account in the fitting procedure of the
effective interactions treated in this paper. In this respect, the
effect can also be relevant for the evaluation of masses with astrophysical
purposes: in a recent fit of the Skyrme interaction \cite{Tondeur.00}
to the known nuclear masses the rotational energy correction is included
(in an approximate way to be discussed below) for deformed nuclei
in the spirit of the PAV and therefore is disregarded for spherical
nuclei. However, taking into account the results just discussed for
$^{208}$Pb the rotational energy correction should also be considered
for spherical nuclei although the procedure to evaluation this quantity
can be much more cumbersome than for the situation of a well deformed
ground state. 

The important quantity in the description of the ground state in the
context of the PBV method is the ground state energy gain due to the
projection and given for even-even nuclei by
\begin{equation}
E_{REC}=E_{HFB}-E^{I=0}
\label{eq:REC}
\end{equation}
It is also called the Rotational Energy Correction (REC) and is a
quantity which increases with increasing deformation of the system
and its typical values are in the range of a few MeV.
This is an important correction when shape coexistence is present
in the system as it can substantially modify the picture emerging
from the mean field and make an excited minimum or shoulder the ground
state. It also reduces in a few MeV the fission barrier heights with
respect to the mean field result and can have important effects even
for spherical nuclei. Several examples will be discussed below in 
Fig.~\ref{fig:EROT}. 

\subsection{An Approximate Evaluation of the Projected Energies: The Strong Deformation
Limit, the Kamlah Expansion and the Like}

\label{sub:ApproxE}
\begin{figure}
\begin{center}\includegraphics[%
  width=1.0\textwidth,
  keepaspectratio]{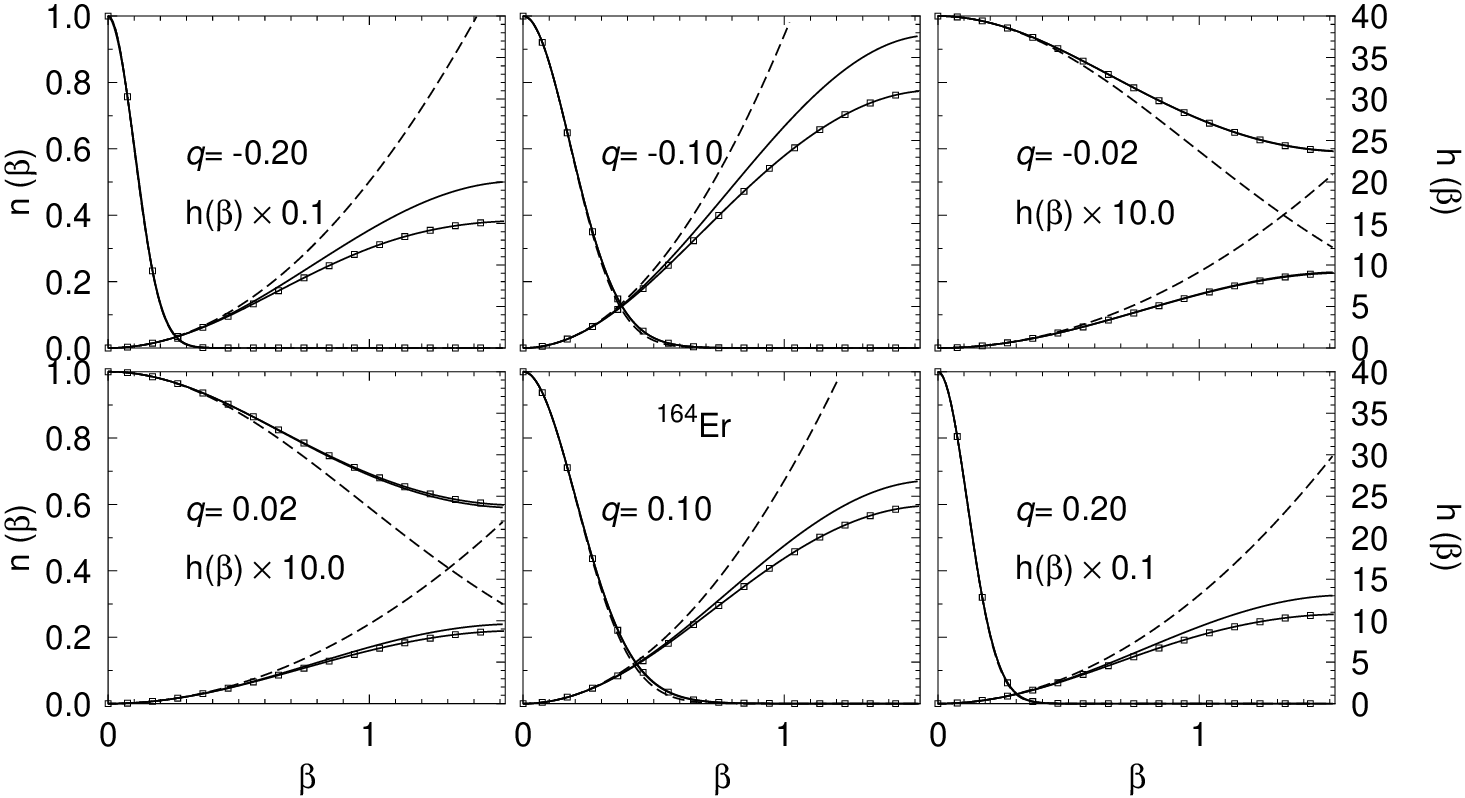}\end{center}

\caption{\label{fig:hbeta}A comparison of the exact $n(\beta)$ and $h(\beta)$
(full line with symbols) with the approximated expressions of 
(\ref{eq:nbetaR}) and (\ref{eq:hbetaR})  (full line) as well as the 
standard ones of (\ref{eq:nbeta})and (\ref{eq:hbeta}) (dashed lines)
for different deformation parameters q.  The quantity $h(\beta)$ has been multiplied 
in each panel by the indicated factor}
\end{figure}

The exact expressions for the projected energy
and transition probabilities are too complicated to have an intuitive
understanding of their effects. Therefore it is convenient to have
estimations given in terms of simple quantities expressible in terms
of the underlying intrinsic state. On the other hand, the evaluation
of the exact projected quantities is rather costly in terms of computational
time an is desirable to have good estimations that are easier to compute.
Such estimations can be easily obtained by means of the Kamlah expansion
\cite{Kamlah,Beck.70} which is valid in the strong deformation limit.
To understand the essence of the approximation let us consider for
simplicity the axially symmetric case and angular momentum zero. The
projected energy is given in this situation by\[
E^{I=0}=\frac{\int_{0}^{\pi/2}d\beta\sin(\beta)h(\beta)n(\beta)}{\int_{0}^{\pi/2}d\beta\sin(\beta)n(\beta)}\]
where, as compared to (\ref{eq:EIAX}) we have reduced the integration
interval to $\pi/2$ by using the signature as a symmetry of the intrinsic
wave function. If the intrinsic wave function is strongly deformed,
the overlap $n(\beta)$ will only be significantly different from
zero when $\beta\approx0$ and will experience a fast decay away from
that value. In fact,  $n(\beta)$ can be approximated by \cite{Ring_Schuck.80}
\begin{equation}
n(\beta)\approx e^{-\frac{1}{2}\langle J_{y}^{2}\rangle\beta^{2}}\,.\label{eq:nbeta}\end{equation}
Next, we assume that $h(\beta)$ is a smooth function of $\beta$
that can be expanded to a good accuracy in a power series up to second
order in $\beta$ \begin{equation}
h(\beta)\approx h_{0}-\frac{1}{2}h_{2}\beta^{2}+\ldots\label{eq:hbeta}\end{equation}
with $h_{0}=\langle H\rangle$ and $h_{2}=-\frac{d²h(\beta)}{d\beta²}_{|\beta=0}$.
Then the projected energy is given by \[
E^{I=0}=h_{0}-\frac{1}{2}h_{2}\Lambda_{0}(\langle J_{y}^{2}\rangle)\]
 where $\Lambda_{0}(\langle J_{y}^{2}\rangle)=\frac{\int_{0}^{\pi/2}d\beta\sin(\beta)\beta^{2}e^{-\frac{1}{2}\langle J_{y}^{2}\rangle\beta^{2}}}{\int_{0}^{\pi/2}d\beta\sin(\beta)e^{-\frac{1}{2}\langle J_{y}^{2}\rangle\beta^{2}}}$.
As this function goes to one when $\langle J_{y}^{2}\rangle>>1$ we
finally arrive to the well known formula for the rotational energy
correction in the strong deformation limit 
\begin{equation}
\label{eq:rotcorr}
E^{I=0}=h_{0}-\frac{1}{2}h_{2}
\end{equation}
which is nowadays widely used in many calculations to get an estimation
of the effect of angular momentum projection in the $I=0$ ground
state. To test the accuracy of this approximation we have computed
for the nucleus $^{164}$Er the exact quantities $n(\beta)$ and $h(\beta)$
for different quadrupole deformations and the results are presented
in Fig. \ref{fig:hbeta} along with the approximate estimates of  (\ref{eq:nbeta})
and (\ref{eq:hbeta}). 

We observe how for strong and moderate deformations the approximated
estimates do rather well and we can conclude that the approximation
to the $I=0$ projected energy is reasonable. However, for small deformations
this is not the case and we expect a failure of the method. As it
was suggested by Reinhard \cite{Reinhard.78,Reinhard.87,Hagino.03} the behavior
of the overlaps can be better approximated for the small deformation
case by using the ansatz \begin{equation}
n(\beta)\approx e^{-\frac{1}{2}\langle J_{y}^{2}\rangle\sin^{2}(\beta)}\label{eq:nbetaR}\end{equation}
and\begin{equation}
h(\beta)\approx h_{0}-\frac{1}{2}h_{2}\sin^{2}(\beta)+\ldots\label{eq:hbetaR}\end{equation}
instead of (\ref{eq:nbeta}) and (\ref{eq:hbeta}). This ansatz is
inspired by the properties $n(\beta)=n(\pi-\beta)$ and $h(\beta)=h(\pi-\beta)$
that the exact quantities have to satisfy when the simplex symmetry
is imposed in the intrinsic wave function and also by analytical results
obtained with harmonic oscillator wave functions. To check the quality
of the new ansatz we have plotted in  Fig. \ref{fig:hbeta} the new quantities. 
We observe how the quality of the ansatz is excellent for small deformations
and it also improves the results at moderate and strong deformations.

\begin{figure}
\begin{center}\includegraphics[%
  width=0.45\textwidth,
  keepaspectratio]{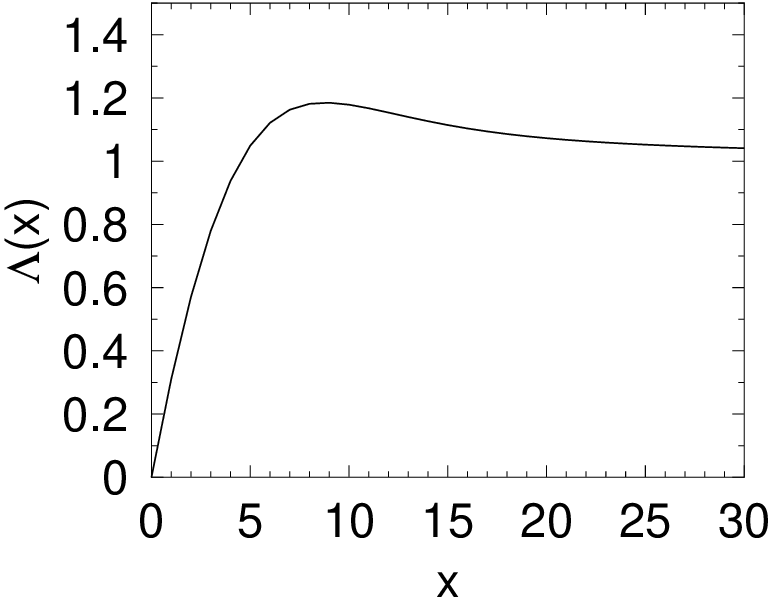}\hspace*{0.5cm}\includegraphics[%
  width=0.45\textwidth,
  keepaspectratio]{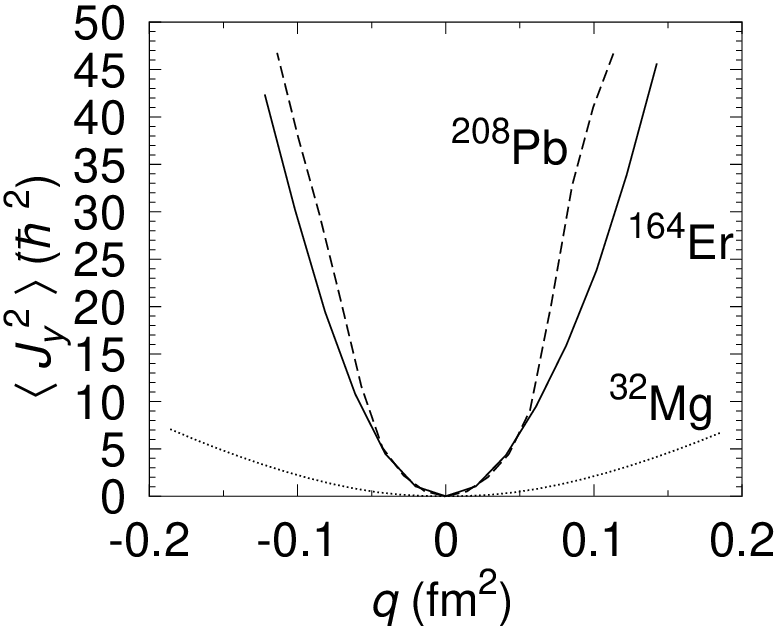}\end{center}

\caption{\label{fig:Lambda}Left panel: The function $\Lambda(x)$ defined in (\ref{eq:Lambda})
as a function of $x$. Right panel: The expectation value 
$\langle J_{y}^{2}\rangle$ as a function of the deformation parameter q}
\end{figure}

Using the new ansatz, the projected energy for $I=0$ is now given
by \begin{equation}
E^{I=0}=h_{0}-\frac{1}{2}h_{2}\Lambda(\langle J_{y}^{2}\rangle)\label{eq:rotcorrR}\end{equation}
with \begin{equation}
\Lambda(\langle J_{y}^{2}\rangle)=\frac{\int_{0}^{\pi/2}d\beta\sin(\beta)\beta^{2}e^{-\frac{1}{2}\langle J_{y}^{2}\rangle\sin^{2}(\beta)}}{\int_{0}^{\pi/2}d\beta\sin(\beta)e^{-\frac{1}{2}\langle J_{y}^{2}\rangle\sin^{2}(\beta)}}.\label{eq:Lambda}\end{equation}
To understand the changes induced by the new ansatz the universal
function $\Lambda(x)$ has been plotted in the left panel of Fig.~\ref{fig:Lambda}
as a function of $x$. There we observe how for big values of $x$ it goes to $1$ 
recovering the standard results and it rapidly decreases for $x$ going to zero
in good agreement with the fact that $\langle J_{y}^{2}\rangle=0$
corresponds (for axially symmetric systems)
to an spherical wave function where the effect of angular
momentum projection is null. Interestingly, the function $\Lambda(x)$
has a maximum at $x\approx7$ thus enhancing the rotational energy
correction at this point. From the shape of this curve it is clear that 
$\Lambda(x)$  induces strong changes in the energy for x values smaller than 7 
and an  smooth modulation of it at larger values. The interesting question 
is to know the deformation range for which x, i.e., $\langle J_{y}^{2}\rangle$,
satisfies $0 \le  \langle J_{y}^{2}\rangle \le 7$. In the right panel of
Fig.~\ref{fig:Lambda} we have represented $\langle J_{y}^{2}\rangle $ 
versus $q$ for the indicated nuclei. We find that the relevant intervals
are $-0.2 \le q \le 0.2$ for  $^{32}$Mg and   $-0.05 \le q \le 0.05$ for
$^{164}$Er and $^{208}$Pb. 

\begin{figure}
\begin{center}\includegraphics[%
  clip,
  width=0.33\textwidth,
  keepaspectratio]{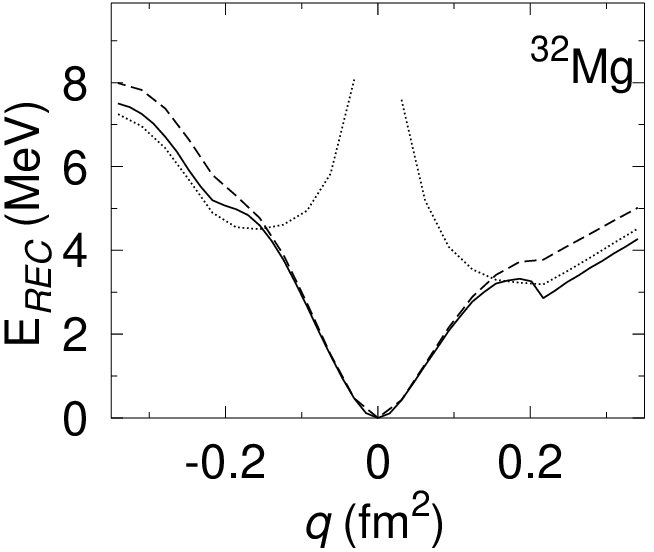}\hspace*{0.1cm}\includegraphics[%
  clip,
  width=0.33\textwidth,
  keepaspectratio]{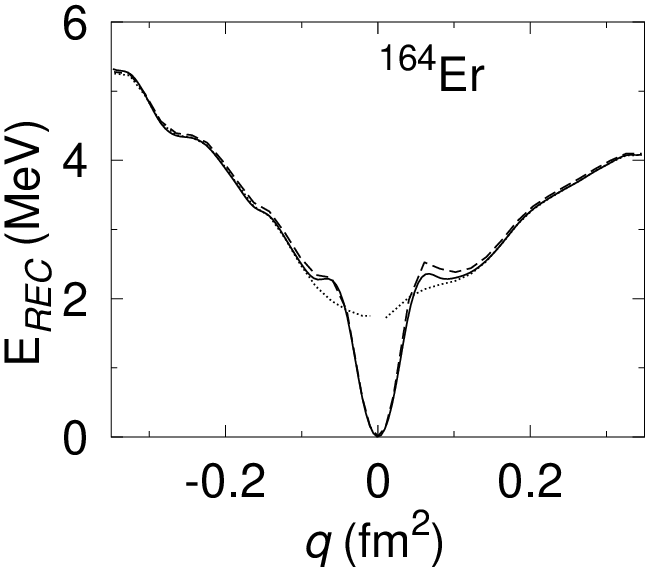}\hspace*{0.1cm}\includegraphics[%
  clip,
  width=0.33\textwidth,
  keepaspectratio]{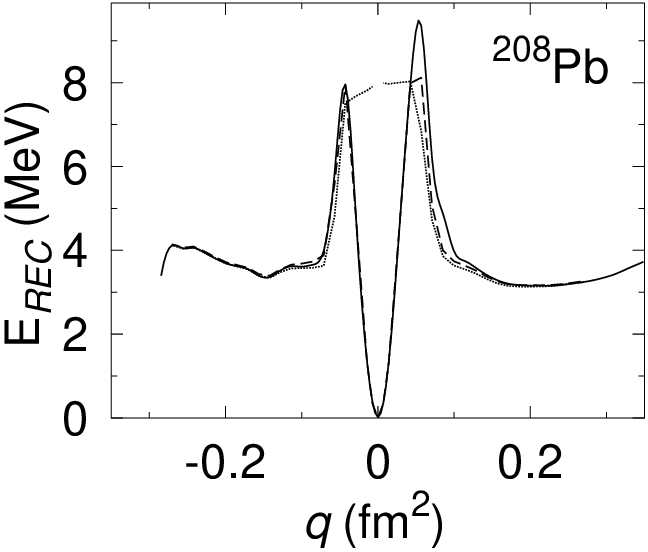}\end{center}

\caption{\label{fig:EROT}Rotational energy corrections as a function of the
quadrupole deformation parameter $q=q_{20}/A^{5/3}$ for $^{32}$Mg, $^{164}$Er
and $^{208}$Pb. The full line represents the exact result of 
(\ref{eq:REC}) whereas the dashed line stands for the approximation
(\ref{eq:rotcorrR}) and the dotted one for the standard correction, see
(\ref{eq:rotcorr})}
\end{figure}

In Fig. \ref{fig:EROT} we now compare the exact rotational energy correction 
(\ref{eq:REC}) at zero spin with the standard approximation (\ref{eq:rotcorr}) 
and the one by Reinhard (\ref{eq:rotcorrR}) for the  mentioned nuclei.
The exact REC  is zero for spherical intrinsic states, it typically 
increases rather abruptly for small deformations and at some point it slides
down and from there on its growing rate stabilizes to a smaller value.
It is interesting to see that the largest changes in the REC are given in
the predicted range $-0.2 \le q \le 0.2$ for  $^{32}$Mg and 
 $-0.05 \le q \le 0.05$ for  $^{164}$Er and $^{208}$Pb.
 As can be observed in this plot the standard 
approximate expression of (\ref{eq:rotcorr}) does pretty bad in the 
corresponding deformation ranges indicated above while  Reinhard's one
does pretty well even in the critical regions. This is an
encouraging result as it can enormously simplify the estimation of
the rotational energy correction in all kind of nuclei, spherical
or deformed, and therefore can make possible its use in massive mass
evaluations like the ones needed for astrophysical purposes.

What we have said up to now is based in the assumption that the intrinsic
wave function is axially symmetric and we have restricted the discussion
to zero spin. In the general case things can be worked out for the
full projection \cite{Kamlah,Beck.70} and the following expression
is obtained in the strong deformation limit\begin{eqnarray}
E^{I} & = & \langle H\rangle-\frac{\langle\Delta\vec{J}^{2}\rangle}{2\mathbb{\mathcal{J}}_{Y}}+\omega\left(\sqrt{I(I+1)-\langle J_{z}^{2}\rangle}-\langle J_{x}\rangle\right)\nonumber \\
 & + & \frac{1}{2\mathbb{\mathcal{J}}_{Y}}\left(\sqrt{I(I+1)-\langle J_{z}^{2}\rangle}-\langle J_{x}\rangle\right)^{2}\label{eq:Eapprox}\end{eqnarray}
where $\mathbb{\mathcal{J}}_{Y}$ is the Yoccoz moment of inertia
given by \begin{equation}
\frac{1}{2\mathbb{\mathcal{J}}_{Y}}=\frac{\langle\Delta H\Delta\vec{J}^{2}\rangle}{2\left(\langle\Delta J_{x}^{2}\rangle^{2}+\langle J_{y}^{2}\rangle^{2}+\langle J_{z}^{2}\rangle^{2}\right)}\label{eq:Yoccoz}\end{equation}
and\begin{equation}
\omega=\frac{\langle H\Delta J_{x}\rangle}{\langle\Delta J_{x}^{2}\rangle}\,.\label{eq:omega}\end{equation}
In order to pin down the physics behind this expression let us assume
that the intrinsic wave function is time reversal invariant. In this
case $\omega=\langle J_{x}\rangle=0$. If, additionally, we impose
an axially symmetric wave function then $\langle J_{z}^{2}\rangle=0$
and the approximate projected energy becomes\[
E^{I}=\langle H\rangle-\frac{\langle\Delta\vec{J}^{2}\rangle}{2\mathbb{\mathcal{J}}_{Y}}+\frac{I(I+1)}{2\mathbb{\mathcal{J}}_{Y}}\]
which corresponds to the energy of a rotor with intrinsic energy $\langle H\rangle-\frac{\langle\Delta\vec{J}^{2}\rangle}{2\mathbb{\mathcal{J}}_{Y}}$,
angular momentum $I$ and moment of inertia $\mathbb{\mathcal{J}}_{Y}$.
In other words, we have a rotational band with the typical $I(I+1)$
behavior and with the Yoccoz moment of inertia. The intrinsic energy
is not simply $\langle H\rangle$ but it is reduced by the so called
rotational energy correction (REC) given by $\frac{\langle\Delta\vec{J}^{2}\rangle}{2\mathbb{\mathcal{J}}_{Y}}$.
Unfortunately, in this case the ansatz proposed by Reinhard is not
easy to generalize and work is needed in order to get equivalent results
in the most general case. In this respected, perhaps the representation
of \cite{Kerman.77} can prove to be useful.

Another interesting aspect of the approximate energy obtained in (\ref{eq:Eapprox})
concerns its use in the context of the Projection Before Variation
method. If one assumes that the approximate energy is a sound approximation
to the projected energy then, requiring the intrinsic state to lead
to a minimum of that expression, is an approximate PBV minimization
process. The minimization process gets rather cumbersome due to the
rotational energy correction term but it has been shown in many examples
that this correction stays rather constant in the domain of validity
of (\ref{eq:Eapprox}) and therefore it is safe to neglect it in the
variational process. Neglecting the variation of the REC one ends
up with the variational equation of the self-consistent cranking model\begin{equation}
\delta\langle\varphi|H-\omega J_{x}|\varphi\rangle=0\label{eq:cranking}\end{equation}
where $\omega$ is determined as to satisfy the constraint\[
\langle\varphi|J_{x}|\varphi\rangle=\sqrt{I(I+1)-\langle J_{z}^{2}\rangle}\,.\]
This equation has to be solved for each value of $I$ and the intrinsic
wave function is therefore $I$ dependent. Another interesting aspect
is related to the fact that $J_{x}$ is odd under time reversal and
therefore the intrinsic wave function breaks time reversal invariance
as well as axial symmetry. The energy spectrum obtained is well represented
by the energy of a rotor\[
E^{I}=\frac{I(I+1)}{2\mathcal{J}_{SCC}(I)}\]
where the self-consistent cranking moment of inertia eventually depends
on $I$. If we now compare this spectrum with the one given by (\ref{eq:Eapprox})
corresponding to the PAV method we notice that the two moments of
inertia are formally different. In the cranking model we have the
self-consistent moment of inertia whereas in the PAV method we have
the Yoccoz one. How different they are depends on the interaction
(probably on the effective mass but there is no formal proof) but
to give a taste of the kind of differences one can expect we can say
that in the case of $^{164}$Er the energy difference $E^{I=2}-E^{I=0}$
is 139.4 keV for the exact PAV whereas the cranking model gives 78.8
keV, that is the cranking result is roughly a factor 0.6 smaller than
the PAV one. Formally, the cranking results are better than the PAV
(even though they can compare worse with experiment due to the interaction
used) as the former are based on the PBV method. However, this does
not mean that we have also to use the self-consistent cranking moment
of inertia (or its close cousin, the Thouless Valatin moment of inertia)
in the evaluation of the rotational energy correction $E_{REC}=\frac{\langle\Delta\vec{J}^{2}\rangle}{2\mathbb{\mathcal{J}}_{Y}}$
as it is sometimes done: the REC has always to be evaluated with the
Yoccoz moment of inertia. The above discussion is also in good agreement
with the results of \cite{Friedman.70,Villars.71} where also an approximate
PBV was carried out but in a different context.

\section{Configuration mixing}

\label{sec:GCM} In the previous sections we have had the opportunity
to glance to several Angular Momentum Projected Energy Surfaces (AMPES)
belonging to different nuclei. In many of them we have obtained several
minima which are close in energy and are also separated by low and
narrow barriers and therefore it is expected that configurations mixing
of those states will lead to a further reduction of the energy of
the states. In this section we will investigate this possibility in
the framework of the Angular Momentum Projected Generator Coordinate
Method (AMPGCM). A general AMPGCM wave function $\mid\Phi_{IM}(\sigma)\rangle$
is written as a linear combination of the {}``generating'' functions
$\mid\Psi_{IM}(\mathbf{q})\rangle$ which are obtained by projecting
onto good angular momentum a set of intrinsic wave functions which
are characterized by several parameter (usually multipole moments
of the mass distribution ) $\mathbf{q=\{}q_{1},q_{2},...\}$. Explicitly
we have\begin{equation}
\mid\Phi_{IM}(\sigma)\rangle=\int d\mathbf{q}f^{I,\sigma}(\mathbf{q})\mid\Psi_{IM}(\mathbf{q})\rangle=\sum_{K}\int d\mathbf{q}\overline{f}_{K}^{I,\sigma}(\mathbf{q})P_{MK}^{I}|\varphi(\mathbf{q})\rangle\,.\label{GCM_ANSATZ}\end{equation}
 The amplitudes $f^{I,\sigma}(\mathbf{q})$ are solutions of the Hill-Wheller
(HW) equation \cite{Hill.53} \[
\sum_{K'}\int d\mathbf{q}'\overline{f}_{K'}^{I,\sigma}(\mathbf{q'})\left(\langle\varphi(\mathbf{q})|HP_{KK'}^{I}|\varphi(\mathbf{q}')\rangle-E^{I,\sigma}\langle\varphi(\mathbf{q})|P_{KK'}^{I}|\varphi(\mathbf{q}')\rangle\right)=0\,.\]
This equation is derived  by imposing the AMPGCM energy to be a minimum. 
As it was mentioned before the calculations are usually restricted to axially
symmetric ($K=0$) configurations and in this case the angular momentum
projected Hill-Wheeler equation reduces to \[
\int d\mathbf{q}'f^{I,\sigma}(\mathbf{q'})\left(\langle\varphi(\mathbf{q})|HP_{00}^{I}|\varphi(\mathbf{q}')\rangle-E^{I,\sigma}\langle\varphi(\mathbf{q})|P_{00}^{I}|\varphi(\mathbf{q}')\rangle\right)=0\,.\]
This is an integral equation where the Hamiltonian kernel $\mathcal{H}^{I}(\mathbf{q},\mathbf{q}')=\langle\varphi(\mathbf{q})|HP_{00}^{I}|\varphi(\mathbf{q}')\rangle$
and the norm overlap $\mathcal{N}^{I}(\mathbf{q},\mathbf{q}')=\langle\varphi(\mathbf{q})|P_{00}^{I}|\varphi(\mathbf{q}')\rangle$
play the central role. The HW equation is a non-orthogonal eigenvalue
equation that is usually recasted in terms of orthogonal quantities
by diagonalizing first the norm overlap\[
\int d\mathbf{q}'\mathcal{N}^{I}(\mathbf{q},\mathbf{q}')u_{l}(\mathbf{q}')=n_{l}^{I}u_{l}(\mathbf{q})\]
what allows to write the norm overlap as\[
\mathcal{N}^{I}(\mathbf{q},\mathbf{q}')=\sum_{l}n_{l}^{I}u_{l}^{*}(\mathbf{q})u_{l}(\mathbf{q'})\,.\]
Inserting this expression into the HW equation and defining the amplitudes\[
g_{l}^{I,\sigma}=(n_{l}^{I})^{1/2}\int d\mathbf{q'}f^{I,\sigma}(\mathbf{q'})u_{l}(\mathbf{q'})\]
and using the inverse relation \begin{equation}
f^{I,\sigma}(\mathbf{q'})=\sum_{l}\frac{g_{l}^{I,\sigma}}{(n_{l}^{I})^{1/2}}u_{l}^{*}(\mathbf{q'})\label{eq:fg}\end{equation}
we end up with the HW equation written as\[
\sum_{l'}(H_{ll'}^{C}-E^{I,\sigma}\delta_{ll'})g_{l'}^{I,\sigma}=0\]
where the collective image of the Hamiltonian kernel $H_{ll'}^{C}$ is
defined as\[
H_{ll'}^{C}=\int d\mathbf{q}d\mathbf{q'}\frac{u_{l}^{*}(\mathbf{q})}{(n_{l}^{I})^{1/2}}\mathcal{H}{}^{I}(\mathbf{q},\mathbf{q}')\frac{u_{l'}(\mathbf{q}')}{(n_{l'}^{I})^{1/2}}\,.\]
The solution of this reduced equation produces the eigenvalues $E^{I,\sigma}$
labeled by the $\sigma$ index and the eigenvectors $g_{l}^{I,\sigma}$
from which we can compute the amplitudes $f^{I,\sigma}(\mathbf{q'})$
using (\ref{eq:fg}). 

In order to interpret the GCM results it is customary to introduce
the collective wave functions \[
g^{I,\sigma}(\mathbf{q})=\int d\mathbf{q}'\left(\mathcal{N}^{1/2}\right)^{I}(\mathbf{q},\mathbf{q}')f^{I,\sigma}(\mathbf{q}')=\sum_{l}g_{l}^{I,\sigma}u_{l}(\mathbf{q})\]
and the reason is that they are orthogonal, contrary to their counterpart
$f^{I,\sigma}(\mathbf{q'})$, and therefore their modulus squared
can be interpreted as a probability amplitude.

In our particular case the set of labels $\mathbf{q}$ reduces to
the quadrupole moment $q_{20}$. If we take into account the axial
symmetry imposed to the intrinsic wave functions and write explicitly
the projector operator we end up with the following expressions for
the norm overlap\begin{equation}
\mathcal{N}^{I}(q_{20},q_{20}^{'})=\left(2I+1\right)\int_{0}^{\frac{\pi}{2}}d\beta\sin(\beta)d_{00}^{I*}(\beta)\langle\varphi(q_{20})\mid e^{-i\beta\hat{J}_{y}}\mid\varphi(q_{20}^{'})\rangle\end{equation}
and the Hamiltonian kernel

\begin{equation}
\mathcal{H}^{I}(q_{20},q_{20}^{'})=\left(2I+1\right)\int_{0}^{\frac{\pi}{2}}d\beta\sin(\beta)d_{00}^{I*}(\beta)\langle\varphi(q_{20})\mid\hat{H}_{DD}e^{-i\beta\hat{J}_{y}}\mid\varphi(q_{20}^{'})\rangle\label{hamiltonian_kernel}\end{equation}
where $\hat{H}_{DD}$ now depends on the density $\overline{\rho}_{\beta}^{GCM}(\vec{r})$
given by

\begin{equation}
\overline{\rho}_{\beta}^{GCM}(\vec{R})=\frac{\langle\varphi(q_{20})\mid\hat{\rho}(\vec{R})e^{-i\beta\hat{J}_{y}}\mid\varphi(q_{20}^{'})\rangle}{\langle\varphi(q_{20})\mid e^{-i\beta\hat{J}_{y}}\mid\varphi(q_{20}^{'})\rangle}\,.\end{equation}
 This is the obvious generalization in the framework of the configuration
mixing calculation of the mixed density prescription described in
Sect. \ref{sub:DD} for the density dependent part of the interaction. 

Finally, from the knowledge of the amplitudes $f^{I,\sigma}(q_{20})$,
we can compute the reduced $B(E2)$ transition probabilities and the
spectroscopic quadrupole moments $Q^{\mathrm{spec}}(I,\sigma)$. This
is one of the main motivations for carrying out a configuration mixing
calculation of angular momentum projected wave functions in the case
of Gogny  and Skyrme forces, since both interactions allow the use
of full configuration spaces and then one is able to compute transition
probabilities and spectroscopic quadrupole moments without effective
charges. In the framework of the AMPGCM the $B(E2)$ transition probability
between the states $(I_{i},\sigma_{i})$ and $(I_{f},\sigma_{f})$
is expressed as (see Sect. \ref{sub:TransProb} for more details) 

\begin{eqnarray}
B(E2,I_{i}\sigma_{i}\rightarrow I_{f}\sigma_{f}) & = & \frac{e^{2}}{2I_{i}+1}\label{BE2_GCM_AM}\\
 & \times & \left|\int dq_{i}dq_{f}f^{*I_{f},{\sigma}_{f}}(q_{f})\langle I_{f}q_{f}\mid\mid\hat{Q}_{2}\mid\mid I_{i}q_{i}\rangle f^{I_{i},{\sigma}_{i}}(q_{i})\right|^{2}\nonumber \end{eqnarray}
 and the spectroscopic quadrupole moment for the state $(I\geq2,\sigma)$
is given by

\begin{eqnarray}
Q\mathrm{^{spec}}(I,\sigma) & = & e\sqrt{\frac{16\pi}{5}}\left(\begin{array}{lcc}
I & 2 & I\\
I & 0 & -I\end{array}\right)\label{ESPECT_GCM_AM}\\
 & \times & \int dq_{i}dq_{f}f^{*I,\sigma}(q_{f})\langle Iq_{f}\mid\mid\hat{Q}_{2}\mid\mid Iq_{i}\rangle f^{I,\sigma}(q_{i})\,.\nonumber \end{eqnarray}

\section{Results}

\label{sec:Results}In order to illustrate the whole procedure we
will discuss now the $^{32}$Mg nucleus in detail and at the end we
will give an overview of the available results obtained with the Skyrme
and Gogny interactions. The nucleus $^{32}$Mg is a nice example as
it is one of the neutron rich magic (N=20) nuclei where the shell
closure is broken. The most convincing experimental evidence for a
deformed ground state in the N=20 isotopes is found in the $^{32}$Mg
nucleus where both the excitation energies of the lowest lying $2^{+}$
\cite{32Mg.E2+} and $4^{+}$ \cite{32Mg.E4+,Guillemaud.02} states
and the $B(E2,0^{+}\rightarrow2^{+})$ transition probability \cite{32Mg.BE2}
have been measured. The low excitation energy of the $2^{+}$ state,
the high value of the $B(E2)$ transition probability and also the
ratio $E(4_{1}^{+})/E(2_{1}^{+})=2.6$ are fairly compatible with
the expectations for a rotational band.

\begin{figure}
\begin{center}\includegraphics[%
  clip,
  width=0.90\textwidth]{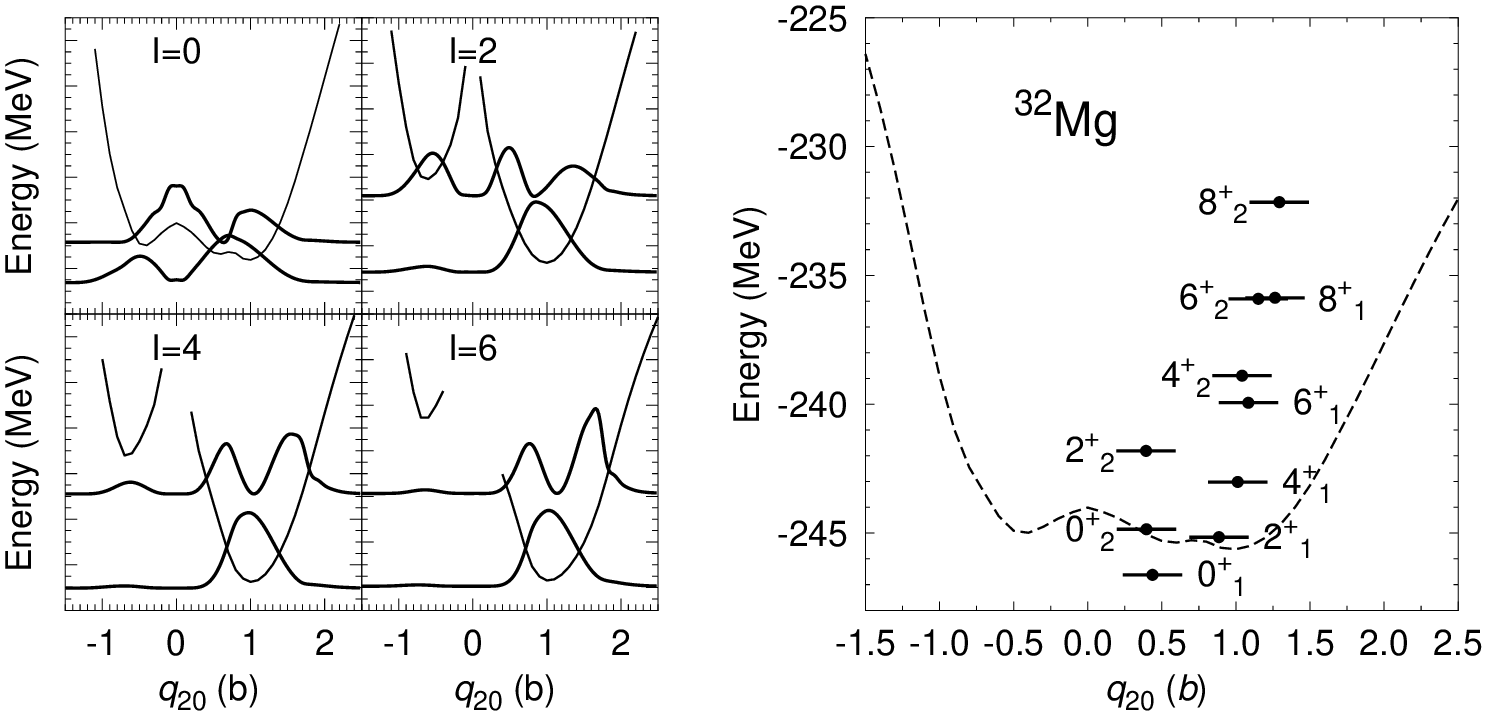} \end{center}

\caption{\label{fig:32Mg}On the left hand side, the square of the {}``collective
wave functions\char`\"{} (in fact, $E^{I,\sigma}+25|g^{I,\sigma}(q_{20})|^{2}$)
for $\sigma=1$ and 2 (thick lines) are plotted along with the corresponding
projected energy (thin line) as a function of $q_{20}$. On the right
hand side, the projected energies $E^{I,\sigma}$ along with the $I=0$
projected energy are plotted as a function of $q_{20}$. The projected
energies have been placed according to their average intrinsic quadrupole
deformation $(\overline{q}_{20})_{\sigma}^{I}$}
\end{figure}

We have carried out Angular Momentum Projected configuration mixing
calculations along the lines described in Sects. \ref{sec:MF}, \ref{sec:AMP}
and \ref{sec:GCM}. The AMP results have been already discussed in 
subsect.~\ref{sub:PBV} and the AMPGCM ones are summarized in Fig.~\ref{fig:32Mg}.
On the left hand side of the figure we have plotted, as a function
of $q_{20}$ and for $I=0$, 2, 4 and 6, and for practical reasons
 a quantity related to the
square of the {}``collective wave functions\char`\"{}, namely 
$E_{\sigma}^{I}+25|g^{I,\sigma}(q_{20})|^{2}$,
for $\sigma=1$ and 2 as well as the projected energy curve. By looking
at the tails of these quantities we can compare the AMPGCM energies
$E_{\sigma}^{I}$ with the corresponding projected energy and have
a feeling of the correlation energy gained by configuration mixing.
For $I=0$ we observe how the ground state has an important mixing
between the prolate and oblate minima. For $I=2$, 4 and 6 the lowest
lying solutions ($\sigma=1$) are well localized inside the prolate
wells whereas the excited solutions ($\sigma=2$) show collective
wave functions reminiscent of a $\beta$ vibrational state. On the
right hand side of Fig. \ref{fig:32Mg} we have plotted the AMPGCM
energies for $\sigma=1$ and 2 along with the $I=0$ projected energy.
The AMPGCM energies have been placed along the $q_{20}$ axis according
to their {}``intrinsic average quadrupole moment\char`\"{} which
is defined as the average quadrupole moment weighted with the {}``collective
wave functions\char`\"{} \begin{equation}
(\overline{q}_{20})_{\sigma}^{I}=\int dq_{20}q_{20}|g^{I,\sigma}(q_{20})|^{2}\,.\end{equation}
 As a result of the mixing with the oblate configurations, the $0_{1}^{+}$
state has a lower intrinsic quadrupole deformation than the minimum
of the $I=0$ projected energy but remains strongly deformed indicating
that the N=20 shell closure is broken in this nucleus. The intrinsic
quadrupole deformation of the $2_{1}^{+}$, $4_{1}^{+}$ and $6_{1}^{+}$
states remains similar the the one of the corresponding minima of
the AMP energy curves as a consequence of the localization inside
the prolate wells of their their collective wave functions. The $4_{2}^{+}$,
$6_{2}^{+}$, etc, states have intrinsic quadrupole deformations similar
to the $4_{1}^{+}$, $6_{1}^{+}$ ones in good agreement with its
quadrupole vibrational character. Another interesting result is the
low excitation energy of the $0_{2}^{+}$ state which is related to
the strong prolate-oblate mixing.

The results we have obtained with the AMPGCM for the $2^{+}$ excitation
energy and $B(E2,0_{1}^{+}\rightarrow2_{1}^{+})$ transition probability
are 1.46 MeV and 395 $e^{2}$ fm$^{4}$, respectively. These results
have to be compared with the angular momentum projected results obtained
with the approximate PBV calculation discussed in Sect. \ref{sub:PBV}
(0.87 MeV and 593 $e^{2}$ fm$^{4}$) and the experiment, 0.88 MeV
and 454$\pm$ 78 $e^{2}$ fm$^{4}$. As a consequence of configuration
mixing the $2^{+}$ excitation energy increases substantially as compared
with the AMP result and differs considerably from the experimental
number. However, the AMPGCM $B(E2)$ transition probability gets reduced
with respect to the AMP result and gets closer to the experimental
value. 
\begin{figure}
\begin{center}\includegraphics[%
  width=0.99\textwidth]{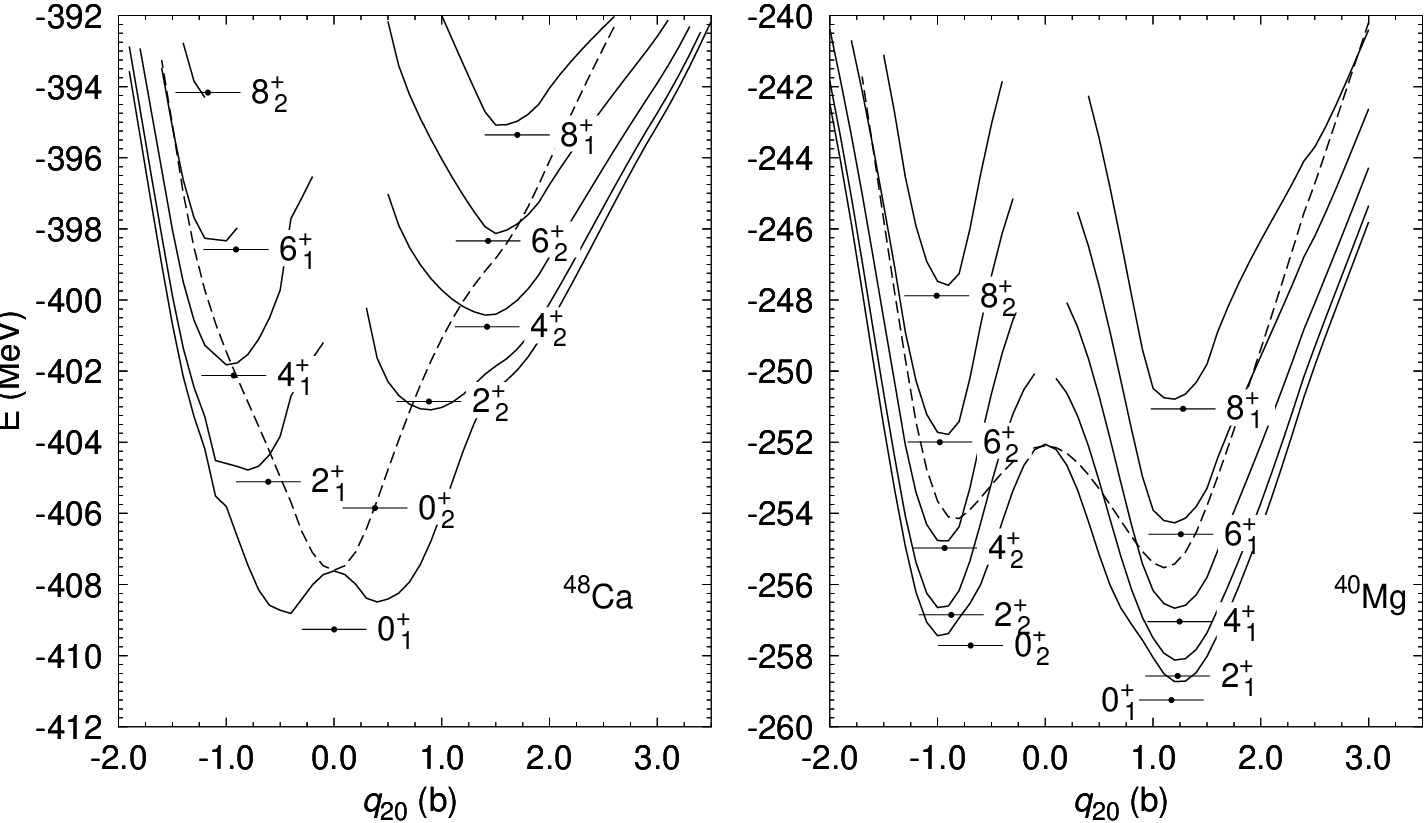}\end{center}

\caption{\label{fig:PES40Mg_48Ca}The PES results for  $^{48}$Ca and  $^{40}$Mg
for $I=0,2,4,6$ and 8, along with the HFB energy (dashed curves)
 as well as the AMPGCM results 
(see comments on the right panel of Fig. \ref{fig:32Mg} for an explanation)}
\end{figure}

$^{32}$Mg  though an spherical nucleus in the mean field approximation
represents the class of nucleus with a second minimum or a shoulder
at an excitation energy comparable with the energy gain by AMP. A different
case is provided by the nucleus  $^{208}$Pb already discussed or $^{48}$Ca, 
see left panel of Fig.~\ref{fig:PES40Mg_48Ca}, which is also spherical at 
the mean field 
level but without any shoulder at relevant. The effect 
of the AMP in this case is to produce two almost degenerated prolate and oblate
minima at small $q$ values.  The behavior of $\langle J^2_y \rangle$ determine in part 
the energy  gain of the ground state in the AMP, see Fig.~\ref{fig:Lambda} and the 
corresponding discussion, which is of the order of 3 MeV for $^{208}$Pb and 
of 1 MeV for  $^{48}$Ca. In this case a configuration mixing calculation
is called for which produces a ground state wave function which is a linear 
combination with similar weights of both minima and on the average the ground
state becomes again spherical, see Fig.~\ref{fig:PES40Mg_48Ca}. An additional 
energy gain (around 0.5 MeV in the case of  $^{48}$Ca) is also provided.
For angular momentum 2, 4, 6 and 8 we find in Fig.~\ref{fig:PES40Mg_48Ca}
that the prolate and oblate minima of the PES of  $^{48}$Ca take place
at larger deformations with increasing spins. This produces larger
barriers between the minima and as a consequence the mixing get smaller
with growing spins and the wave functions remain well localized in the
corresponding minima. To complete our discussion we have plotted the
mean field results, the PES and the AMPGCM energy levels in the right
panel for the {\em well deformed} nucleus $^{40}$Mg, though the REC's are 
similar for both plotted nuclei, the effect of the projection on the
mean field energy is quite different. The MF prolate ground state and
the secondary  oblate minima,  after the AMP remain approximately
at the same $q_{20}$ values and the same relative energy. The energy gain
of the minima amounts to 3.5 MeV. The structure of the MFA energy surface is 
also roughly maintained by the PES at higher spins. The configuration mixing
calculations provide different results than in the spherical nucleus,
one obtain two rather well defined rotational bands, i.e., with rather
constant intrinsic quadrupole moment, and less mixing in the wave functions.
The energy gain of the ground state by the configuration mixing amounts  
about 600 keV.


In the following we will present results obtained with the Skyrme
and the Gogny interaction. The calculations with the Skyrme interaction
have been carried out mostly with the SLy4 \cite{SLy4} or the SLy6
\cite{SLy6} parameterizations of the force. In the AMPGCM ansatz 
a projection onto good particle number is also performed.
In the case of the Gogny interaction the parameterization D1S \cite{Berger.84}
has been used. A simultaneous particle number projection has not been carried out 
because the finite range of the force  considerably increases the computational
burden as compared to the Skyrme interaction. As the field is pretty
new there are not too many results and in some cases the nuclei computed
with the Skyrme interaction are the same as the ones computed with
the Gogny force. In those situations we will concentrate mainly in
the Gogny results as the Skyrme ones are mostly in qualitative (and
in some cases, quantitative) agreement with the former interaction.
Most of the calculations have been focusing in light nuclei either
with or without an excess of neutrons. Also calculations in heavier
nuclei are available in the neutron deficient Lead isotopes although
some more results will come in the near future. 

We will start first commenting on the Skyrme results: the well deformed
$^{24}$Mg nucleus was discussed in \cite{Valor.00a} with the SLy4
force and zero range pairing interactions (depending and not depending
on the density) and a reasonable agreement was obtained for the ground
state rotational band although the band was too stretched. The intra-band
$B(E2)$ transition probabilities were well reproduced. In \cite{Bender.03}
the doubly magic $^{16}$O was studied with the SLy4 parameterization
and focusing on the deformed $0^{+}$ excited states and the corresponding
rotational bands built on top of them. A rather good agreement was
found between the calculations and the experimental results. The structure
of the first excited $0^{+}$ state was discussed in terms of multi
particle-hole components and it was found that the structure of this
state was a 4p-4h excitation in agreement with previous shell model
explanations. It is interesting to point out that the ground state
energy was lowered by 2.3 MeV with respect to the mean field as a
consequence of the correlations induced by the AMPGCM method. The
neutron deficient $^{186}$Pb, which is a very nice example of shape
coexistence with and spherical ground state and two excited states
below 1 MeV of prolate and oblate character respectively, was studied
in \cite{Duguet.03} with the SLy6 parameterization and zero range
density dependent pairing interaction. The excitation energy of the
prolate $0^{+}$excited state was very well reproduced but this was
not the case for the energy of the nearby oblate $0^{+}$state. The
rotational band built on top of the prolate $0^{+}$ excited state
came up too spread out as compared with the experimental data. The
energies of both $0^{+}$ states are pushed up to higher energies (1.05
MeV and 1.39 MeV for the prolate and oblate states, respectively)
when the SLy4 force is used worsening the agreement with experiment.
Finally, in \cite{Bender.03-2} the nuclei $^{32}$S, $^{36}$Ar,
$^{38}$Ar and $^{40}$Ca were studied. The interest was here the
normal and super-deformed bands recently found in those nuclei. The
normal deformed and super-deformed band heads come up at a reasonable
excitation energy in all the nuclei but the rotational bands built
on top of them come always too spread out as compared with the experiment.
However, the $B(E2)$ values are in rather good agreement with experiment.
The existence of SD bands in those nuclei is connected to a partial
occupation of the $f_{7/2}$ sub-shell.

The calculations with the Gogny force have focused in neutron rich
light nuclei in order to investigate the erosion of the $N=20$ and
$N=28$ shell closures. To study the $N=20$ region we computed 
\cite{Rayner.2000} the Angular Momentum Projected energy curves 
for $^{30-34}$Mg and $^{32-38}$Si and from the minima of the projected curves 
we determined
the lowest state for each angular momentum. In this way we were able
to demonstrate that the ground state of the magic $^{32}$Mg was deformed
and traced back the breaking of the shell closure to the partial occupancy
of the $f_{7/2}$ intruder orbital. We also obtained a deformed ground
state in $^{34}$Mg and for the other nuclei we concluded that they
were showing shape coexistence. Concerning the excitation energy of
the $2^{+}$states we got a rather nice agreement with experiment
as well as for the $B(E2)$ transition probabilities. From our preliminary
analysis it was clear that a proper description of those nuclei should
involve configuration mixing in the context of the AMPGCM. This calculation
was carried out for  $^{30-24}$Mg \cite{Rayner.2000b} and  
for $^{32-36}$Si \cite{Rayner.2001}. As a general rule, the inclusion of configuration
mixing increases the energy of the $2^{+}$states as compared with
the AMP results worsening the agreement with experiment but the $B(E2)$
get usually reduced improving the agreement with experiment. It was
established the spherical character of $^{30}$Mg and the deformed
one of $^{32-34}$Mg as well as the spherical character of the ground
states of all the Si isotopes studied.

To analyze the erosion of the $N=28$ shell closure we have considered 
\cite{Rayner.2002a} the isotones from $^{40}$Mg up to $^{48}$Ca
as well as the Sulfur isotopes $^{38-42}$S. It is found
that the drip line nucleus $^{40}$Mg is prolate deformed breaking
thereby the $N=28$ shell closure, the nucleus $^{42}$Si is also
deformed but in this case in the oblate side. The nuclei $^{44}$S
and $^{46}$Ar are found to show shape coexistence whereas $^{48}$Ca
is found to be spherical as expected. Concerning the Sulfur isotopes
they are found to be to a greater or lesser extent prolate deformed.
The excitation energies of the $2^{+}$states and specially their
transition probabilities to the ground state are rather well reproduced
(the energies always come up rather high as compared with the experiment).
The two neutron and two protons separation energies are also well
reproduced and it is found that the effect of AMPGCM is only significant
in a few cases while in the other the results are pretty close to
the HFB ones.
\begin{figure}
\begin{center}\includegraphics[%
  width=0.80\textwidth]{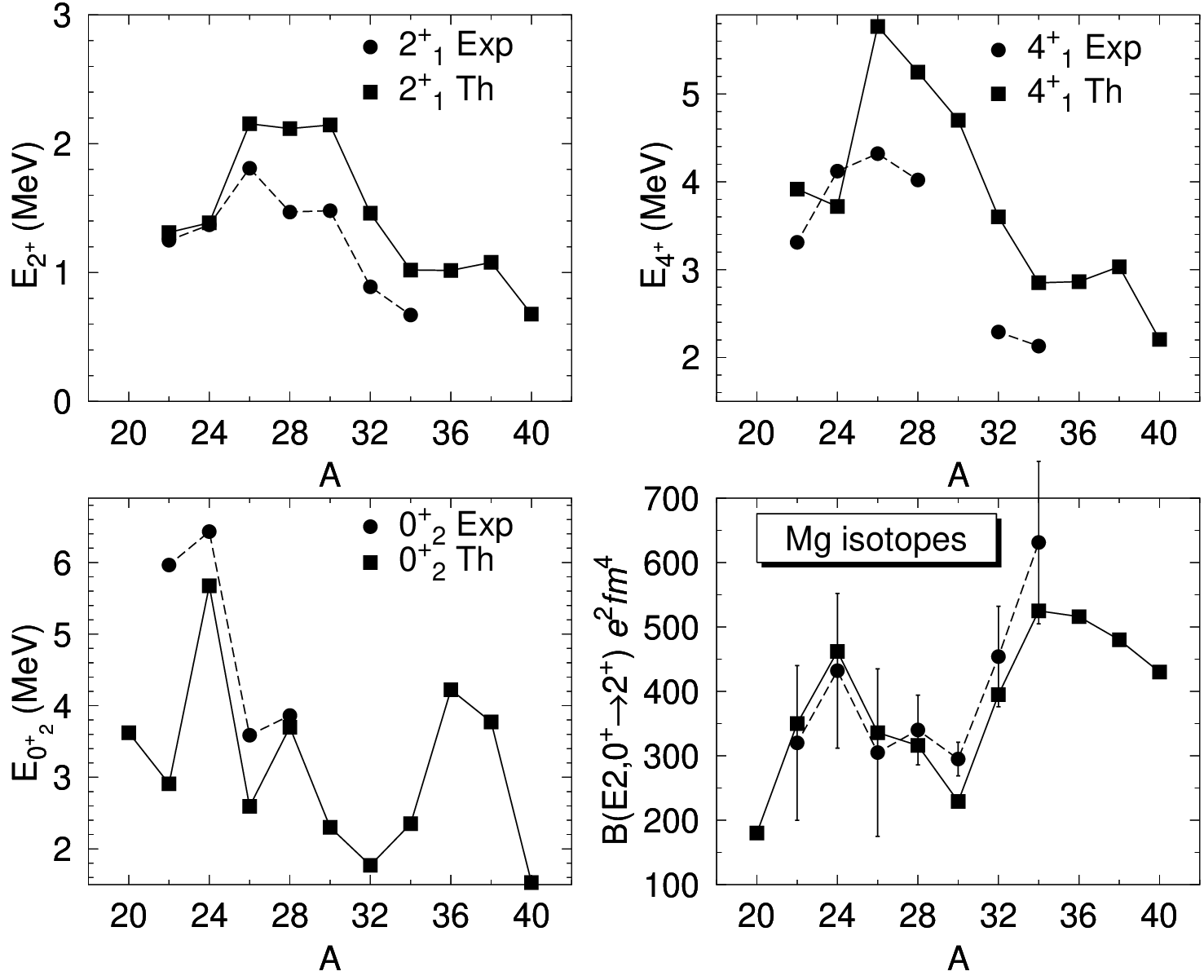}\end{center}

\caption{\label{fig:Mg} Comparison of the theoretical results (full line)
for the $2^{+}$, $4^{+}$, and $0_{2}^{+}$ excitation energies and
$B(E2,0^{+}\rightarrow2^{+})$ transition probabilities for the Mg
isotopic chain with the experimental data (dashed line)}
\end{figure}

The Magnesium isotopic chain from A=20 until the neutron drip line
A=40 was studied in \cite{Rayner.2002b}. The results obtained for
the $2^{+}$, $4^{+}$and $0_{2}^{+}$ excitation energies as well
as the $B(E2,0^{+}\rightarrow2^{+})$ are represented in Fig. \ref{fig:Mg}
for the whole chain and compared with the experimental data. We observe
that the isotopic trend is quite well reproduced but the $2^{+}$and
$4^{+}$ excitation energies come up too high whereas the $0_{2}^{+}$ come
up too low. Interestingly we are able to reproduce the sliding down
of the $0_{2}^{+}$excitation energy in going from $^{24}$Mg to $^{26}$Mg
and due to the fact that whereas $^{24}$Mg is a well deformed nucleus
$^{26}$Mg is a shape coexistent one \cite{Rayner.03a}. One of the
main results of our calculation is that the nuclei from $^{32}$Mg
to $^{40}$Mg show a prolate deformed ground state. To finish the
discussion of the Figure let us finally comment that the experimental
$B(E2)$ transition probabilities are pretty well reproduced by our
calculations.

\begin{figure}
\begin{center}\includegraphics[%
  width=0.50\textwidth]{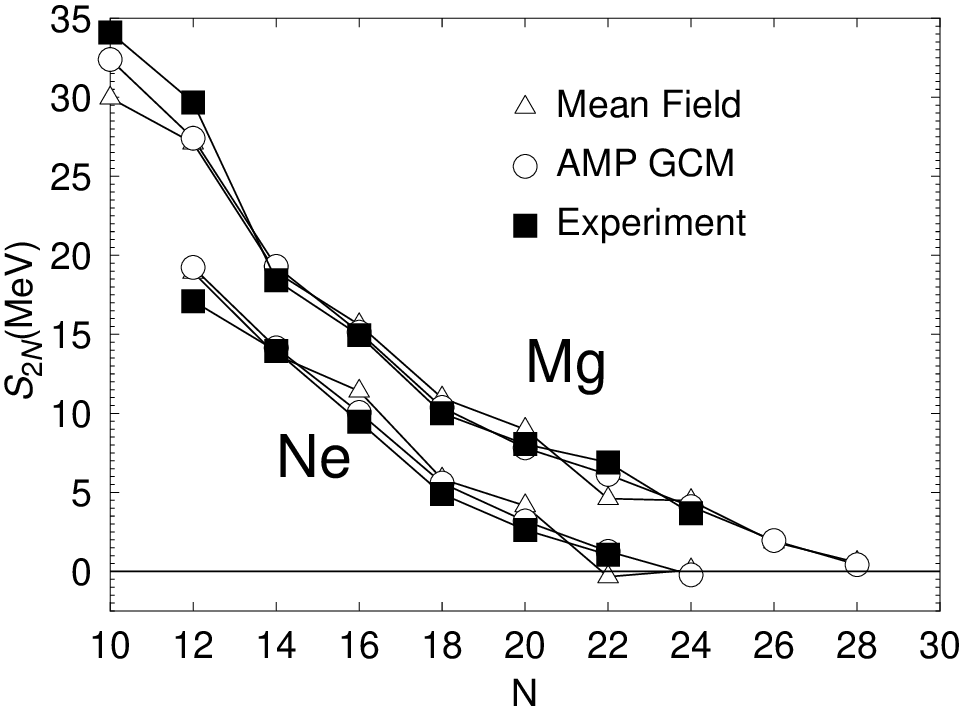}\end{center}

\caption{\label{fig:MgS2N} Two neutron separation energies $S_{2N}$ as a
function of neutron number for the isotopic Mg and Ne chains. Results
for the mean field and the AMPGCM calculations are present along with
the experimental data}
\end{figure}

The AMPGCM method not only provides the excitation energy of the states
but also the ground state energy and therefore one can compute two
neutron or two proton separation energies and compare them with the
HFB results. This comparison is made in Fig. \ref{fig:MgS2N} for
the Mg \cite{Rayner.2002b} and Ne \cite{Rayner.03} isotopic chains
and there we observe how the AMPGCM method improves the agreement
with experiment as compared with the HFB results. It is particularly
interesting the $N=22$ result that is strongly related to the erosion
of the $N=20$ shell closure, the inclusion of correlations that made
$^{32}$Mg deformed are the responsible for the better agreement of
the $S_{2N}$ separation energy in this case. Finally, let us mention
that according to our calculations the neutron drip line for the Mg
isotopes is located at $N=28$. For the Ne isotopes the AMPGCM predicts
(contrary to the mean field) an stable $^{32}$Ne in good agreement
with experiment.

Calculations similar to the ones presented for the Mg isotopes have
also been carried out for the Ne isotopic chain in \cite{Rayner.03}.
The conclusions are similar to the ones found for the Mg chain and
they will not be discussed here. 

Finally, let us comment on the fact that independently of the force
used the rotational bands obtained with the AMPGCM method are too
stretched as compared with the experimental data. On the other hand
the $B(E2)$ transition probabilities compare pretty well with experiment.
This {}``feature'' of the method might have to do with the lack in the present 
formalism of additional correlations like triaxialities, two-quasiparticle
mixing and so on.


\section{Outlook and Future Developments}

\label{sec:Outlook}In this paper we have described how to carry out
Angular Momentum Projection (AMP) with effective interactions of the
Skyrme and Gogny type. We have illustrated the procedure with several
 examples. The main outcome of those calculations is that
AMP strongly modifies the energy landscape as a function of the mass
quadrupole moment of the system and therefore the naive procedure
of Projection After Variation (PAV) (i.e. projecting after the intrinsic
state is determined at the mean field level by optimizing the intrinsic
energy) leads in many cases to the wrong answer. On the other hand,
the restricted Projection Before Variation (PBV) (i.e. the intrinsic
state is determined by optimizing the projected energy) where only
the quadrupole moment is allowed to vary yields much more consistent
results. We have also learned that the projected energy landscapes
have in many cases coexisting minima and therefore it is important
to consider configuration mixing which is implemented in the framework
of the Angular Momentum Projected Generator Coordinate Method (AMPGCM).
The results shown indicate the relevance of configuration mixing in
many cases. We have also discussed approximate ways to compute the
projected quantities and in this context we have compared the present
methodology with the results of the cranking model. All the methods
considered have been illustrated with relevant examples and at the
end we have given a rather exhaustive account of the available theoretical
results. As the field is rather new there are still further improvements
to be made: a) It has to be explored if there are other relevant degrees
of freedom apart from the quadrupole moment that could be used in
the context of the restricted PBV. b) The suitability of the method
for the description of odd nuclei has to be explored. c) The octupole
degree of freedom has to be incorporated in order to extend the present
method into the realm of negative parity states. d) Triaxial and/or
time reversal breaking admixtures have to be incorporated. e) Approximations
to compute the projected quantities have to be explored and their
suitability assessed, and a long etcetera !

\section{Acknowledgements}

\label{sec:Acknowledgements}All the work presented here for the Angular
Momentum Projection plus Generator Coordinate Method with the Gogny force is 
part or an outcome of the Ph. D. thesis work of Rayner Rodríguez-Guzman to 
whom we are very indebted. This work has been supported in part by DGI, 
Ministerio de Ciencia y Tecnología, Spain, under Project BFM2001-0184.


\begin{thebibliography}{10}
\bibitem{Bender.review}M. Bender and P.-H. Heenen, Rev. of Mod. Phys. 75, 121 (2003)
\bibitem{Mang.75}H.J. Mang, Phys. Rep. 18, 325 (1975)
\bibitem{Ring_Schuck.80}P. Ring and P. Schuck, \emph{The Nuclear Many Body Problem} (Springer,
Berlin, 1980)
\bibitem{Blaizot.85}J.-P. Blaizot and G. Ripka, \emph{Quantum theory of Finite systems}
(MIT, Cambridge, MA, 1985)
\bibitem{Peierls.57}R.E. Peierls and J. Yoccoz, Proc. Phys. Soc. London \textbf{A}70,
381 (1957)
\bibitem{Hill.53}D.L.Hill and J.A.Wheeler, Phys. Rev. 89, 1102 (1953)
\bibitem{Brown.01}B.A. Brown, Prog. Part. Nucl. Phys. 47, 517 (2001)
\bibitem{Caurier.98}E. Caurier, F. Nowacki, A. Poves and J. Retamosa, Phys. Rev. \textbf{C}58,
2033 (1998); A. Schmidt, I. Schneider, C. Friessner, A.F. Lisetskiy,
N. Pietralla, T. Sebe, T. Otsuka and P. von Brentano, Phys. Rev. \textbf{C}62,
044319 (2000)
\bibitem{Schmid.01}K.W. Schmid, Prog. Part. Nucl. Phys. 46, 145 (2001)
\bibitem{Schmid.87}K.W. Schmid and F. Grümmer, Rep. Prog. Phys. 50, 731 (1987)
\bibitem{Hara.95}K. Hara and Y. Sun, Int. J. Mod. Phys. \textbf{E}4, 637 (1995)
\bibitem{Otsuka.01}T. Otsuka, M. Homma, T. Mizusaki, N. Shimizuo and Y. Utsuno, Prog.
Part. Nucl. Phys. 47, 319 (2001)
\bibitem{Skyrme.56}T.H.R. Skyrme, Phil. Mag. 1, 1043 (1956); Nucl. Phys. 9, 615 (1959)
\bibitem{Vautherin.72}D. Vautherin and D.M. Brink, Phys. Rev. \textbf{C}3, 626 (1972)
\bibitem{Engel.75}Y.M. Engel, D.M. Brink K. Goeke, S.J. Krieger and D. Vautherin, Nucl.
Phys. \textbf{A}249, 215 (1975)
\bibitem{Dech_Gogny.80}J. Decharge and D. Gogny, Phys. Rev. \textbf{C}21, 1568 (1980)
\bibitem{SkyrIII}M. Beiner, H. Flocard, Nguyen Van Giai and P. Quentin, Nucl. Phys.
\textbf{A}238, 29 (1975)
\bibitem{SkM}H. Krivine, J. Treiner and O. Bohigas, Nucl. Phys. \textbf{A}366,
155 (1980)
\bibitem{SkM*}J. Bartel, P. Quentin, M. Brack, C. Guet and H.-B Hakansson, Nucl.
Phys. \textbf{A}386, 79 (1982)
\bibitem{SLy4}E. Chabanat, P. Bonche, P. Haensel, J. Meyer and R. Schaeffer, Nucl.
Phys. \textbf{A}627, 710 (1977)
\bibitem{SLy6}E. Chabanat, P. Bonche, P. Haensel, J. Meyer and R. Schaeffer, Nucl.
Phys. \textbf{A}635 (1998) 231; Nucl. Phys. \textbf{A}643, 441 (1998)
\bibitem{Berger.84}J.F. Berger, M. Girod and D. Gogny, Nucl. Phys. \textbf{A}428, 23c
(1984)
\bibitem{GognyD1P}M. Farine, D. Von-Eiff, P. Schuck, J. F. Berger, J. Dechargé and M.
Girod, J. Phys. \textbf{G}: Nucl. Part. Phys. 25, 863 (1999)
\bibitem{Anguiano.01}M. Anguiano, J.L. Egido and L.M. Robledo, Nucl. Phys. \textbf{A}683,
227 (2001)
\bibitem{Robledo.94}L. M. Robledo, Phys. Rev. \textbf{C}50, 2874 (1994); J.L. Egido, L.M.
Robledo and Y. Sun, Nucl. Phys. \textbf{A}560, 253 (1993)
\bibitem{Varsha.88}D. A. Varshalovich, A. N. Moskalev and V. K Khersonskii, \emph{Quantum
Theory of Angular Momentum} (World Scientific, Singapore, 1988)
\bibitem{Corbett.71}J.O. Corbett, Nucl. Phys. \textbf{A}169, 426 (1971)
\bibitem{Kerman.77}A.K. Kerman and N. Onishi, Nucl. Phys. \textbf{A}281, 373 (1977)
\bibitem{Balian.69}R.Balian and E.Brézin, \emph{Il Nuovo Cimento}, vol. LXIV \textbf{B},
37 (1969)
\bibitem{Neergard.83}K.Neergard and E.Wust, Nucl. Phys. \textbf{A}402, 311 (1983)
\bibitem{Bonche.91}P. Bonche, J. Dobaczewski, H. Flocard, P.-H. Heenen, Nucl. Phys. \textbf{A}530,
149 (1991)
\bibitem{Hagino.03} K. Hagino, G.B. Bertsch and P.-G. Reinhard, 
Phys. Rev. \textbf{C}68, 024306 (2003)
\bibitem{Duguet.03B}T. Duguet and P. Bonche, Phys. Rev. \textbf{C}67, 054308 (2003)
\bibitem{Bonche.90}P. Bonche, J. Dobaczewski, H. Flocard, P.-H. Heenen and J. Meyer,
Nucl. Phys. \textbf{A}510, 466 (1990)
\bibitem{Valor.00}A. Valor, J.L. Egido and L.M. Robledo, Nucl. Phys. \textbf{A}665,
46 (2000)
\bibitem{Rayner.2002b}R. Rodr\'{\i}guez-Guzm\'{a}n, J.L. Egido and L.M. Robledo, Nucl.
Phys. \textbf{A}709, 201 (2002)
\bibitem{Tondeur.00}F. Tondeur, S. Goriely, J.M. Pearson and M. Onsi, Phys. Rev. \textbf{C}62,
024308 (2000)
\bibitem{Kamlah}A. Kamlah, Z. Physik 216, 52 (1968)
\bibitem{Beck.70}R. Beck, H.J. Mang and P. Ring, Z. Physik 231, 26 (1970)
\bibitem{Reinhard.78}P.-G. Reinhard, Z. Physik \textbf{A}285, 93 (1978)
\bibitem{Reinhard.87}P.-G. Reinhard and K. Goeke, Rep. Prog. Phys. 50, 1 (1987)
\bibitem{Friedman.70}W.A. Friedman and L. Wilets, Phys. Rev. \textbf{C}2, 892 (1970)
\bibitem{Villars.71}F. Villars and N. Schmeing-Rogerson, Annals of Physics 63, 433 (1971)
\bibitem{32Mg.E2+}C. D\'{e}traz, D. Guillemaud, G. Huber, R. Plapisch, M. Langevin,
F. Naulin, C. Thibault, L.C. Catraz and F. Touchard, Phys. Rev. \textbf{C}19
(1979) 164; D.Guillemaud-Mueller, C.D\'{e}traz, M.Langevin, F.Naulin,
M.De Saint-Simon, C.Thibault, F.Touchard and M.Epherre, Nucl. Phys.
\textbf{A}426, 37 (1984)
\bibitem{32Mg.E4+}G. Klotz, P. Baumann, M. Bounajma, A. Huck, A. Knipper, G. Walter,
G. Marguier, C. Richard-Serre, A. Poves and J. Retamosa, Pys. Rev.
\textbf{C}47, 2502 (1993)
\bibitem{Guillemaud.02}D. Guillemaud-Mueller, Eur. Phys. J. \textbf{A} 13, 63 (2002)
\bibitem{32Mg.BE2}T.Motobayashi, Y.Ikeda, Y.Ando, K.Ieki, M.Inoue, N.Iwasa, T.Kikuchi,
M.Kurokawa, S.Moriya, S.Ogawa, H.Murakami, S.Shimoura, Y.Yanagisawa,
T.Nakamura, Y.Watanabe, M.Ishiara, T.Teranishi, H.Okuno and R.F.Casten,
Phys. Lett. \textbf{B}346, 9 (1995)
\bibitem{Valor.00a}A. Valor, P.-H. Heenen and P. Bonche, Nucl. Phys. \textbf{A}671, 145
(2000)
\bibitem{Bender.03}M. Bender and P.-H. Heenen, Nucl.Phys. \textbf{A}713, 390 (2003)
\bibitem{Duguet.03}T. Duguet, M. Bender, P. Bonche and P.-H. Heenen, Phys.Lett. \textbf{B}559,
201 (2003)
\bibitem{Bender.03-2}M. Bender, H. Flocard and P.-H. Heenen, ( \emph{preprint nucl-th/0305021})
\bibitem{Rayner.2000}R. Rodríguez-Guzmán, J.L. Egido and L.M. Robledo, Phys. Lett. \textbf{B}474,
15 (2000)
\bibitem{Rayner.2000b}R. Rodríguez-Guzmán, J.L. Egido and L.M. Robledo, Phys. Rev. \textbf{C}62,
054319 (2000)
\bibitem{Rayner.2001}R. Rodríguez-Guzmán, J.L. Egido and L.M. Robledo, Acta Phys. Pol.
\textbf{B}32, 2385 (2001)
\bibitem{Rayner.2002a}R. Rodríguez-Guzmán, J.L. Egido and L.M. Robledo, Phys. Rev. \textbf{C}65,
024304 (2002)
\bibitem{Rayner.03a}R. Rodr\'{\i}guez-Guzm\'{a}n, J.L. Egido and L.M. Robledo, AIP \emph{Conference
Proceedings} 656, 303 (2003)
\bibitem{Rayner.03}R. Rodr\'{\i}guez-Guzm\'{a}n, J.L. Egido and L.M. Robledo, Eur.
Phys. J. \textbf{A}17, 37 (2003)\end{thebibliography}
\end{document}